\documentclass[usenatbib]{mnras}
\usepackage{graphicx}
\usepackage{latexsym}
\usepackage{color}

\def\ltsima{$\; \buildrel < \over \sim \;$}
\def\simlt{\lower.5ex\hbox{\ltsima}}
\def\gtsima{$\; \buildrel > \over \sim \;$}
\def\simgt{\lower.5ex\hbox{\gtsima}}
\def\kms{{\rm\,km\,s^{-1}}}

\def\AA{$\; \buildrel \circ \over {\rm A}$}

\def\deg{^\circ}
\def\s{\ifmmode \widetilde \else \~\fi}
\def\={\overline}

\def\spose#1{\hbox to 0pt{#1\hss}}

\def\lta{\mathrel{\spose{\lower 3pt\hbox{$\mathchar"218$}}
     \raise 2.0pt\hbox{$\mathchar"13C$}}}
\def\gta{\mathrel{\spose{\lower 3pt\hbox{$\mathchar"218$}}
     \raise 2.0pt\hbox{$\mathchar"13E$}}}
\def\Dt{\spose{\raise 1.5ex\hbox{\hskip3pt$\mathchar"201$}}}    % upper case
\def\dt{\spose{\raise 1.0ex\hbox{\hskip2pt$\mathchar"201$}}}    % lower case

\def\dotsfill{\leaders\hbox to 1em{\hss.\hss}\hfill}

\def\FeH{{\rm[Fe/H]}}

\title[The \emph{Pristine} survey]{The \emph{Pristine} survey I: Mining the Galaxy for the most metal-poor stars\thanks{Based on observational programmes 15AC20, 15AF14, 15AF97, 16AC20, 16AC98, and 16AF14 obtained with MegaPrime/MegaCam, a joint project of CFHT and CEA/DAPNIA, at the Canada-France-Hawaii Telescope (CFHT) which is operated by the National Research Council (NRC) of Canada, the Institut National des Science de l'Univers of the Centre National de la Recherche Scientifique (CNRS) of France, and the University of Hawaii. The observations at the Canada-France-Hawaii Telescope were performed with care and respect from the summit of Maunakea which is a significant cultural and historic site. }} 
\author[E. Starkenburg et al.,]{Else Starkenburg$^1$\thanks{E-mail: estarkenburg@aip.de}, Nicolas Martin$^{2,3}$, Kris Youakim$^1$,
David S. Aguado$^{4,5}$, 
\newauthor Carlos Allende Prieto$^{4,5}$, Anke Arentsen$^1$, Edouard J. Bernard$^6$, Piercarlo  
\newauthor Bonifacio$^7$, Elisabetta Caffau$^7$, Raymond G. Carlberg$^8$, Patrick C\^{o}t\'{e}$^9$,
\newauthor Morgan Fouesneau$^3$, Patrick Fran\c{c}ois$^{7,10}$, Oliver Franke$^{11}$, Jonay I. Gonz\'{a}lez 
\newauthor Hern\'{a}ndez$^{4,5}$, Stephen D. J. Gwyn$^9$, Vanessa Hill$^{12}$, Rodrigo A. Ibata$^2$, Pascale
\newauthor Jablonka$^{7,13}$, Nicolas Longeard$^2$, Alan W. McConnachie$^9$, Julio F. Navarro$^{14}$, 
\newauthor Rub\'en S\'anchez-Janssen$^{9,15}$, Eline Tolstoy$^{16}$, Kim A. Venn$^{14}$ \\
$^1$ Leibniz Institute for Astrophyics Potsdam (AIP), An der Sternwarte 16, D-14482 Potsdam, Germany\\
$^2$ Universit\'e de Strasbourg, CNRS, Observatoire astronomique de Strasbourg, UMR 7550, F-67000 Strasbourg, France\\
$^3$ Max-Planck-Institut f\"{u}r Astronomie, K\"{o}nigstuhl 17, D-69117 Heidelberg, Germany \\
$^4$ Instituto de Astrof\'{i}sica de Canarias, V\'{i}a L\'{a}ctea, 38205 La Laguna, Tenerife, Spain \\
$^5$ Universidad de La Laguna, Departamento de Astrof'sica, 38206 La Laguna, Tenerife, Spain \\
$^6$ Universit\'{e} C\^{o}te d'Azur, OCA, CNRS, Lagrange, France \\
$^7$ GEPI, Observatoire de Paris, PSL Research University, CNRS, Place Jules Janssen, 92190 Meudon, France \\
$^8$ Department of Astronomy \& Astrophysics, University of Toronto, Toronto, ON M5S 3H4, Canada \\
$^9$ NRC Herzberg Astronomy and Astrophysics, 5071 West Saanich Road, Victoria, BC V9E 2E7, Canada \\
$^{10}$ Universit\'{e} de Picardie Jules Verne, 33 rue St-Leu, 80080 Amiens, France \\
$^{11}$ Institut f\"{u}r Physik und Astronomie, Universit\"{a}t Potsdam, Karl-Liebknecht-Str. 24/25, 14476 Golm, Germany\\
$^{12}$ Laboratoire Lagrange, Universit\'e de Nice Sophia-Antipolis, Observatoire de la C\^{o}te d'Azur, CNRS, Bd de l'Observatoire, \\
CS 34229, 06304 Nice cedex 4, France \\
$^{13}$ Laboratoire d'astrophysique, \'{E}cole Polytechnique F\'{e}d\'{e}rale de Lausanne (EPFL), Observatoire, 1290 Versoix, Switzerland \\
$^{14}$ Dept. of Physics and Astronomy, University of Victoria, P.O. Box 3055, STN CSC, Victoria BC V8W 3P6, Canada \\
$^{15}$ Royal Observatory Edinburgh, Blackford Hill, Edinburgh, EH9 3HJ, UK \\
$^{16}$ Kapteyn Astronomical Institute, University of Groningen, Landleven 12, 9747AD Groningen, Netherlands} 

\begin{document}

\maketitle

\begin{abstract}
We present the \emph{Pristine} survey, a new narrow-band photometric survey focused on the metallicity-sensitive Ca H \& K lines and conducted in the northern hemisphere with the wide-field imager MegaCam on the Canada-France-Hawaii Telescope (CFHT). This paper reviews our overall survey strategy and discusses the data processing and metallicity calibration. Additionally we review the application of these data to the main aims of the survey, which are to gather a large sample of the most metal-poor stars in the Galaxy, to further characterise the faintest Milky Way satellites, and to map the (metal-poor) substructure in the Galactic halo. The current \emph{Pristine} footprint comprises over 1,000 deg$^2$ in the Galactic halo ranging from $b \sim 30\deg$ to $\sim 78\deg$ and covers many known stellar substructures. We demonstrate that, for SDSS stellar objects, we can calibrate the photometry at the 0.02-magnitude level. The comparison with existing spectroscopic metallicities from SDSS/SEGUE and LAMOST shows that, when combined with SDSS broad-band $g$ and $i$ photometry, we can use the CaHK photometry to infer photometric metallicities with an accuracy of $\sim $0.2 dex from $\FeH = -0.5$ down to the extremely metal-poor regime ($\FeH < -3.0$). After the removal of various contaminants, we can efficiently select metal-poor stars and build a very complete sample with high purity. The success rate of uncovering $\FeH_\mathrm{SEGUE} < -3.0$ stars among $\FeH_\mathrm{Pristine} < -3.0$ selected stars is 24\% and 85\% of the remaining candidates are still very metal poor (\FeH$ < -2.0$). We further demonstrate that \emph{Pristine} is well suited to identify the very rare and pristine Galactic stars with $\FeH < -4.0$, which can teach us valuable lessons about the early Universe.
\end{abstract}


\section{Introduction}

The subject of research into metal-poor stars is truly unique to our
Local Group. Only here can we resolve individual stars and therefore single out the very rare pristine stars among the much more numerous metal-rich populations. At the same time this topic is instructive for so
many disciplines in astronomy: it guides our understanding of the physics behind star
formation (in particular in the absence of metals), supernovae, the early build-up of galaxies and the epoch of
reionization. From a theoretical perspective, it has
been shown that the very first stars that formed in the Universe are more likely to be very massive due to the limited
cooling processes available to them in the absence of any metals \citep[see][and references therein]{Bromm13,Greif15}. However, it is still heavily debated whether lower mass stars could have also been formed in the fragmentation process. This leaves open the exciting possibility that truly first stars could still be shining in our Galaxy today, waiting to be uncovered.

The number of extremely and ultra metal-poor stars \citep[with $\FeH < -3.0$ and $\FeH < -4.0$, respectively;][]{Beers05} analysed in depth has grown a lot in the last decades thanks to laborious efforts to find and characterise them. There are now $\sim $10 stars known with intrinsic iron-abundances
below $\FeH = -4.5$ \citep{Christlieb02, Frebel05, Norris07, Caffau13a, Hansen14, Keller14, AllendePrieto15, Bonifacio15, AllendePrieto15, Frebel15, Caffau16}\footnote{We dismiss here the class of iron-poor stars that are chemically peculiar post-asymptotic giant branch stars in which many elements like iron have condensed out into grains \citep[see][for an analysis of some of these stars and their relation to intrinsically iron-poor stars]{Waelkens91,Takeda02,Venn08,Venn14}.}. Some of the most iron-poor stars display abundance patterns that could bear the imprint of the explosion physics of first supernovae. For instance, the \emph{SkyMapper} Southern Sky Survey star SMSS J031300.36--670839.3 which is the current record holder most iron-poor star at $\FeH < -7.1$, shows a very high carbon abundance as well as a remarkably high [Mg/Ca] ratio ([Mg/Ca] = 3.1) a combination that, according to some models, links it to the explosion of a $\sim 60$ solar-mass star without any metals \citep{Keller14,Bessell15,Nordlander17}.
 
As another example of the constraining power of new observations in this field, it seemed that, based on the handful of stars found in this regime, all ultra iron-poor stars showed a very high carbon abundance, typically [C/Fe]$ > $1.0, up to [C/Fe]$ > $4.0 \citep[see for instance the compilations in][]{Spite13, Norris13}. It was suggested that such an abundance might be needed to reach the ``metallicity floor'', a necessary amount of metals
available in a gas cloud such that it can cool and form a low-mass star \citep[e.g.,][]{Frebel07}. However, \citet{Caffau11} discovered an ultra metal-poor star that was shown not to be severely carbon-enhanced, impacting our theories on
star formation in the early Universe. This counterexample clearly indicates various formation routes for ultra metal-poor stars, a conclusion that was strengthened very recently in the bulge region of the Galaxy by \citet{Howes15} who analysed a sample of 23 stars below $\FeH = -2.3$ --- including one star at $\FeH = -3.94 \pm 0.09$\ --- and found none of them to be Carbon-enhanced. Nevertheless, it seems that, at least in the halo, a very large fraction of observed stars \citep[$ \sim $32\%,][]{Yong13b} still displays a high carbon abundance \citep[][]{Beers05}. There are some indications for a further dependence of the presence of carbon-enhanced metal-poor stars and their sub-type on environment (with disk height, \citet{Frebel06}; in the inner Galaxy, \citet{Howes15}; in the inner versus outer halo, \citet{Carollo14}; and in the halo versus dwarf galaxies, \citet{Starkenburg13, Skuladottir15}). However, statistics are lacking and a larger sample of these stars is needed across various Galactic environments to truly understand this early epoch of the formation of the Galaxy. 

\begin{figure}
\includegraphics[width=\linewidth]{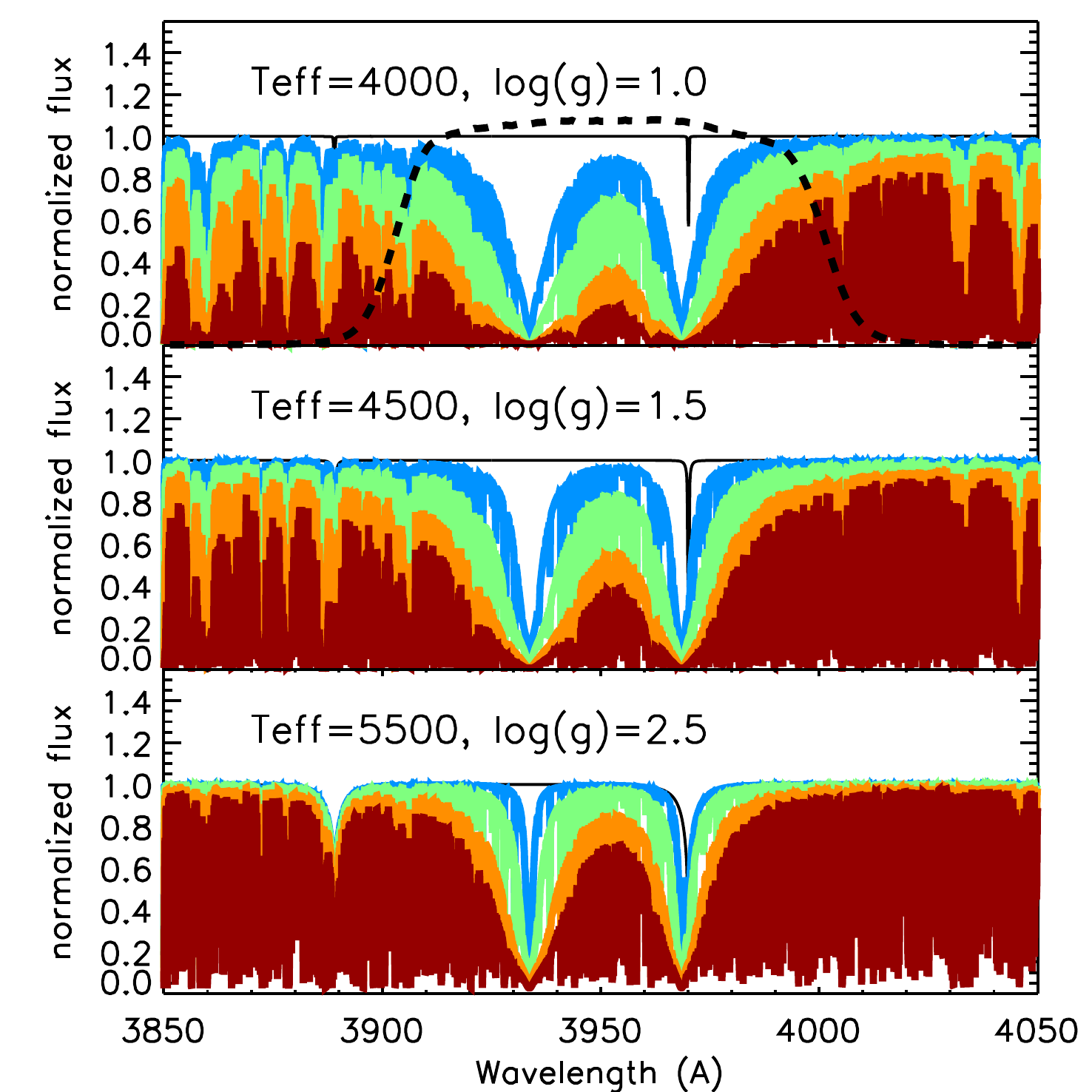}
\caption{Synthetic spectra using MARCS stellar atmospheres and the Turbospectrum code (see section \ref{sec:filter} for details) of stars on three different places on the giant branch with metallicities $\FeH = 0.0$ (red), $\FeH = -1.0$ (orange), $\FeH = -2.0$ (green), $\FeH = -3.0$ (blue), and for a star with no metals (black). In the top panel the throughput of the Ca H \& K filter used in \emph{Pristine} is overplotted (black dashed line).}
\label{fig:CaHK}
\end{figure}

A high fraction of the known extremely metal-poor stars was discovered in the HK and Hamburg/ESO (HES) surveys \citep{Beers85,Christlieb02}. Both these surveys were centred around the very strong Ca H \& K features in the spectrum, using a grism imaging
technique. The HK interference-filter/objective-prism project surveyed 7000 deg$^{2}$ of sky to a B-magnitude of 15.5 and found $\sim $100 stars with estimated metallicity $\FeH < -3.0$. A slightly larger number of such stars were discovered in the extragalactic Southern Sky surveyed by the HES. Interestingly, no stars with $\FeH < -4.0$ were found in the HK survey, whereas several were discovered in the HES. Most likely, this is due to an increased depth in the latter survey (a limiting magnitude of B$ \sim $17--17.5), which allowed the survey to probe more thoroughly into the outer halo component. The target lists from these surveys have been dominating the field of research on pristine stars for many years (e.g., see \citealt{Cohen13} for a recent compilation of results from the targets of the HES survey).

In the coming decade it is expected that the search for metal-poor stars will
intensify and that finally a large sample of these stars will be uncovered,
allowing us to refine our knowledge of the early Universe based on detailed
studies within our own Galaxy. More or less metallicity-blind and sparse but very large spectroscopic
endeavours such as SDSS \citep[Sloan Digital Sky Survey,][]{York00}, SEGUE \citep[Sloan Extension for Galactic Understanding and Exploration,][]{Yanny09,Eisenstein11}, and LAMOST \citep[Large Sky Area Multi-Object Fiber Spectroscopic Telescope,][]{Cui12} with its LEGUE project \citep[LAMOST Experiment for Galactic Understanding and Exploration,][]{Deng12} offer good candidate lists from their
initial low-resolution spectra to follow-up at higher resolution \citep{Caffau12,Aoki13,AllendePrieto15,Aguado16}. Future, ever larger spectroscopic surveys will be carried out by WEAVE \citep[a multi-object survey spectrograph for the 4.2-m William Herschel Telescope,][]{Dalton12,Dalton14}, PFS \citep[Subaru Prime Focus Spectrograph,][]{Takada14}, 4MOST \citep[4-metre Multi-Object Spectroscopic Telescope,][]{deJong14}, or are planned, such as MSE \citep[Maunakea Spectroscopic Explorer,][]{McConnachie16b,McConnachie16a}. Combinations
of various sources of broad-band photometry from large surveys are also a
source of candidates \citep[see][for a technique that combines WISE (Wide-Field Infrared Survey Explorer) infrared
photometric bands and optical bands to search for metal-poor targets among bright stars]{Schlaufman14,Casey15}. 
 
Narrow-band photometric surveys around the Ca H \& K lines have a great potential to provide a breakthrough in the search for metal-poor stars. The clear advantage of narrow-band photometry over spectroscopic methods is its efficiency: all stars in the field of view are measured simultaneously and no pre-selection is needed. Photometry can also handle crowded fields better than objective-prism methods and will reach fainter objects with a similar observing time. Any very metal-poor star shows weaker
Ca H \& K absorption features (see Figure \ref{fig:CaHK}), setting it apart from more metal-rich stars of the same temperature, approximated through the broad-band colour spectrum. In stars with similar broad-band colours one can subsequently compare the relative flux in a narrow-band filter across these strong absorption features and infer the metallicity of the star. The dependence of the strength of the Ca H \& K lines on other stellar parameters such as surface gravity is much weaker than the dependence on either temperature or metallicity and can be ignored as a first order approach, in particular in the metal-poor regime.

\begin{figure*}
\includegraphics[width=\linewidth]{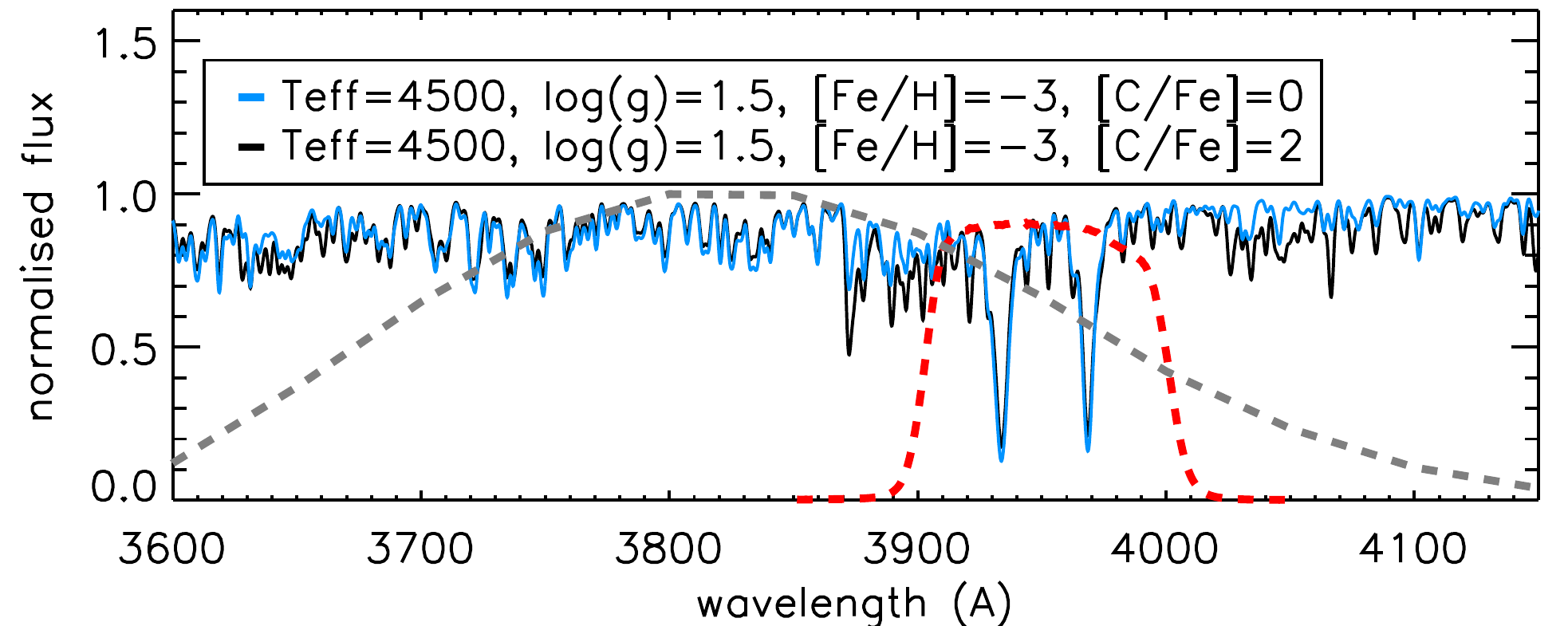}
\caption{Scaled throughput curves of the \emph{Pristine} Ca H \& K filter (red) and the \emph{SkyMapper} $v$ filter (grey) plotted over synthetic model spectra of an extremely metal-poor giant. The black spectrum is additionally enhanced in C and N by 2 dex. \label{fig:filters}}
\end{figure*}

This technique of narrow-band photometry of the Ca H \& K metallicity-dependent features is already conducted on a large scale by the \emph{SkyMapper} survey. This facility is an automated wide-field, 1.35-meter survey telescope at Siding Spring Observatory \citep[e.g.,][]{Keller07,Murphy09,Keller12}. It is designed to map all of the Southern Hemisphere in a set of SDSS-like $ugriz$ filters and, additionally, a $v$ narrow-band filter that includes the Ca H \& K doublet \citep[see][and the dashed grey filter curve in Figure \ref{fig:filters}]{Bessell11}. The \emph{Skymapper} team has been using this filter to search for (extremely) metal-poor stars, their sample of candidates have already revealed some intriguing very metal-poor stars that were subsequently followed up with spectroscopy \citep{Howes14, Keller14, Howes15}. Spectroscopic follow up of several metal-poor stars also selected from photometry with a narrow-band Ca K filter in an area near the Galactic bulge are presented by \citet{Koch16}.

In this paper, we describe the \emph{Pristine} survey, a narrow-band Ca H \& K survey in the Northern Hemisphere. This survey utilises the unique facility of a (novel) Ca H \& K filter for the MegaCam wide-field imager on the 3.6-meter Canada France Hawaii Telescope (CFHT) on the excellent site of Maunakea in Hawaii, in combination with existing broad-band photometry from SDSS. \emph{Pristine} focusses its footprint on high-Galactic-latitude regions ($b > 30\deg$) to remain within the SDSS footprint. There is a wealth of known substructures within the survey regions --- consisting of dwarf galaxies, globular clusters and stellar streams --- which are all very promising structures to hunt for the oldest stars \citep[e.g.,][]{Starkenburg17}. The survey data and the data reduction process, including the photometric calibration, are described in Section \ref{sec:survey}. Our overlap with the SDSS footprint also ensures that we are essentially self-calibrated with the help of the SDSS and SEGUE spectra. Section \ref{sec:segue} shows how well we can separate stars of various metallicities and clean our sample of contaminants. In Section \ref{sec:science} we summarise the main science cases enabled by \emph{Pristine}. We show how metallicity sensitive photometry, as performed by \emph{Pristine}, can probe the Galaxy out to its virial radius. Not only does it allow for an efficient search for ultra metal-poor stars, but it also provides a mapping of the metal-poor (and probably oldest) components of the Milky Way halo that will help dissect the Milky Way's past. 

\section{The survey and data reduction}\label{sec:survey}
\subsection{The Ca H \& K filter properties }\label{sec:filter}

Figures \ref{fig:CaHK} and \ref{fig:filters} illustrate the properties of the Ca H \& K filter used for \emph{Pristine} (a.k.a. CFHT/Megacam narrow-band filter 9303\footnote{See \url{http://www.cfht.hawaii.edu/Instruments/Filters/megaprime.html} for the filter curve.}). The filter is manufactured by Materion and was received by CFHT in November 2014. It is designed to be close to top-hat in its throughput filter curve as a function of wavelength. By design, the filter has a width of $\sim 100$ \AA\ and covers the wavelengths of the Ca H \& K doublet lines (at 3968.5 and 3933.7 \AA), thereby also allowing for a typical spread in radial velocity among the stars observed in the Galactic halo, making it especially suited for our science. For the remainder of the paper we will refer to this filter as the CaHK filter, and to its measured magnitudes as \emph{CaHK} magnitudes. For comparison, we also show the \emph{SkyMapper} $v$ filter used for the same purpose. Clearly, the CFHT CaHK filter is narrower and more top-hat, resulting in a better sensitivity to the Ca H \& K line strength and less danger of leakage from other features such as strong molecular bands in C- and N-enhanced stars, as can be seen from the difference between the blue and black spectra on the figure.

\begin{figure*}
\includegraphics[width=\linewidth]{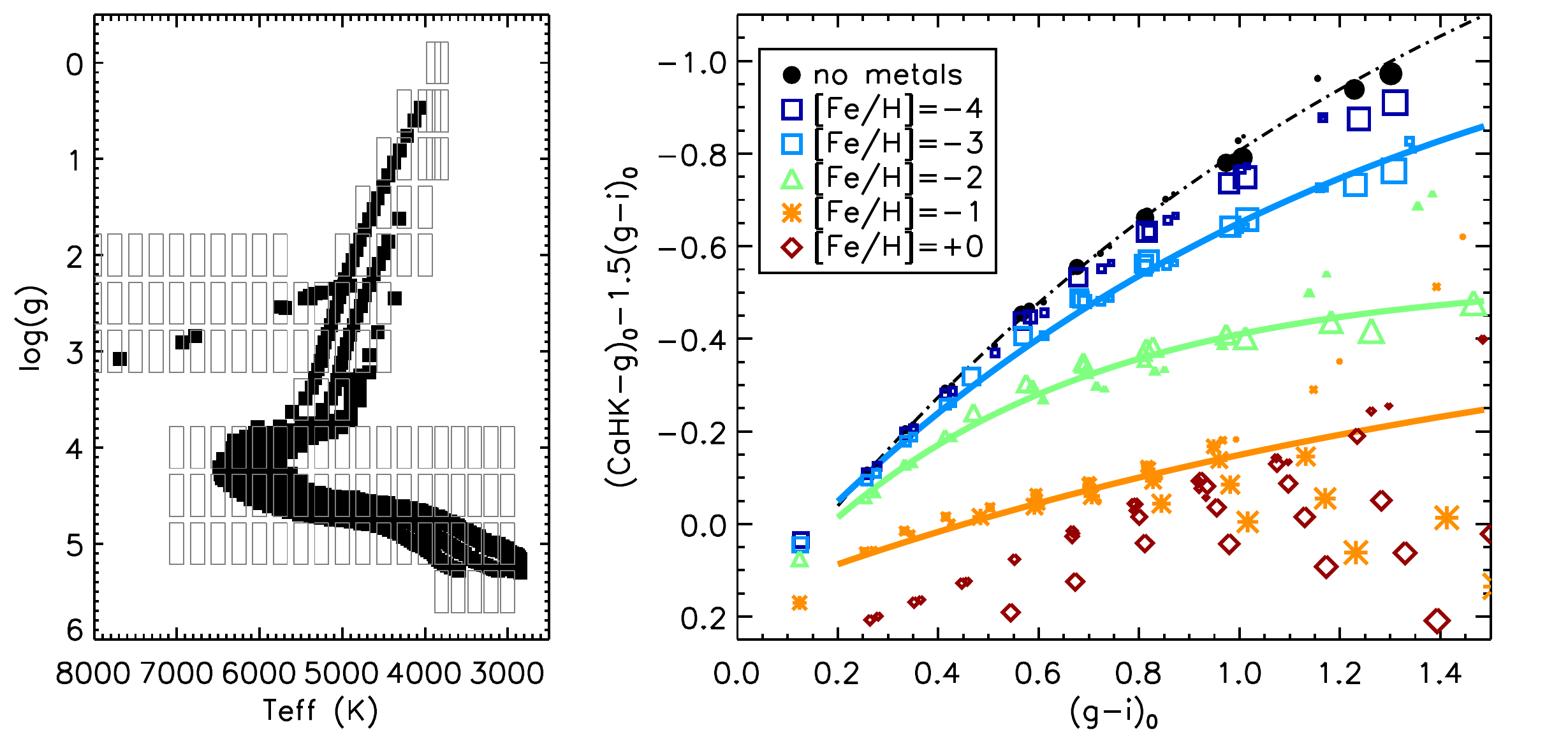}
\caption{Left: Besan\c{c}on model prediction (black points) and the spectral synthesis grid used in these tests (grey) in Teff, log(g) space. Spectra are computed for the full parameter space for [Fe/H]$ = -4.0, -3.0, -2.0, -1.0$ and $+0.0$. Right: All predicted stars are matched to their most representative spectrum in the grid and colours are calculated for the spectra. This panel demonstrates how combinations of broad-band SDSS colours and the new narrow band filter cleanly separate the various metallicity stars in the sample. Additionally, the most metal-poor stellar atmosphere models are run with no metal absorption lines, resulting in the black circles. The coloured lines represent exponential fits to the symbols of metallicities \FeH$ = -1,-2$ and $-3$ and no metals (same colour-coding, redwards of $g-i = 1.0$ the giant branch is followed instead of the main sequence dwarfs). \label{fig:synt}}
\end{figure*}

The expected discriminative power of the CaHK filter is further demonstrated in Figure \ref{fig:synt}. The left panel of this figure shows the range of a spectral library in temperature and gravity parameter space and compares this with the stars as expected in a 100 deg$^{2}$ high-latitude field in an anti-centre direction as indicated by the Besan\c{c}on model of the Galaxy \citep{Robin03}. We have created a library of synthetic spectra, illustrated here by the grey boxes, with large ranges in effective temperature, gravity and metallicity ($-4.0 < $[Fe/H]$ < +0.0$) using MARCS (Model Atmospheres in Radiative and Convective Scheme) stellar atmospheres and the Turbospectrum code \citep{Alvarez98,Gustafsson08,Plez08}. All elements are treated as scaled from solar abundances, with exception of the $\alpha$-elements that are enhanced relative to scaled solar by +0.4 in the models with \FeH$ < -1.0$. Several individual spectra from this library are shown in Figures \ref{fig:CaHK} and \ref{fig:filters}. For each combination of stellar parameters in the synthetic grid, we evaluate if indeed such a star is physically expected, by checking if that box of temperature, log(g) and metallicity is filled with a star in the Besan\c{c}on model. All verified synthetic spectra are subsequently integrated with the response curves of the photometric SDSS bands and average response curve of the CaHK filter. If a star with [Fe/H]$ < -2$ is found for that combination of stellar parameters, we include all models of [Fe/H]$ < -2$ and lower, motivated by the fact that these stars are too rare to find all possible physical combinations in a 100 deg$^{2}$ field-of-view in the Besan\c{c}on model, but that isochrones generally change very little at these lowest metallicities. We additionally synthesise all [Fe/H] = $-4$ models while taking out any absorption lines by atoms or molecules heavier than Li. This set of additional synthetic spectra represents our approximation to stars without any metals at all. The right panel of Figure 3 demonstrates that the CaHK filter in combination with SDSS broad bands is a very powerful tool to select metal-poor stars. The additional $(g-i)_0$ term on the y-axis is purely used to flatten the relation such that the fanning out of the different metallicities is oriented from top to bottom. The size of the symbols is inversely proportional to their surface gravities (larger symbols are giant stars, smaller symbol stars are main-sequence dwarfs). As can be seen from Figure \ref{fig:synt}, the surface gravities have some impact on the exact placement of the star in this colour-colour space but, especially for low-metallicity stars, the effect of gravity differences between a main-sequence star and a red giant is much less pronounced than the metallicity information. The synthetic spectra that are run without any metals are shown as black symbols. As the Ca H \& K absorption lines get weaker with decreasing metallicity, so does the vertical space between the models that are 1 dex apart in \FeH\ space. According to this test, it should nevertheless be possible to distinguish $\FeH = -3$ and $\FeH = -4$ stars with a CaHK filter, in particular at lower temperatures when the lines are stronger. An exponential fit is provided to each of the model metallicities of \FeH$ = -1,-2,-3$ and the no-metals models, shown here as lines colour-coded according to their corresponding metallicity. We compare these predictions to data in Section \ref{sec:segue}.

\subsection{Observations and coverage}
\begin{figure*}
\includegraphics[width=\linewidth]{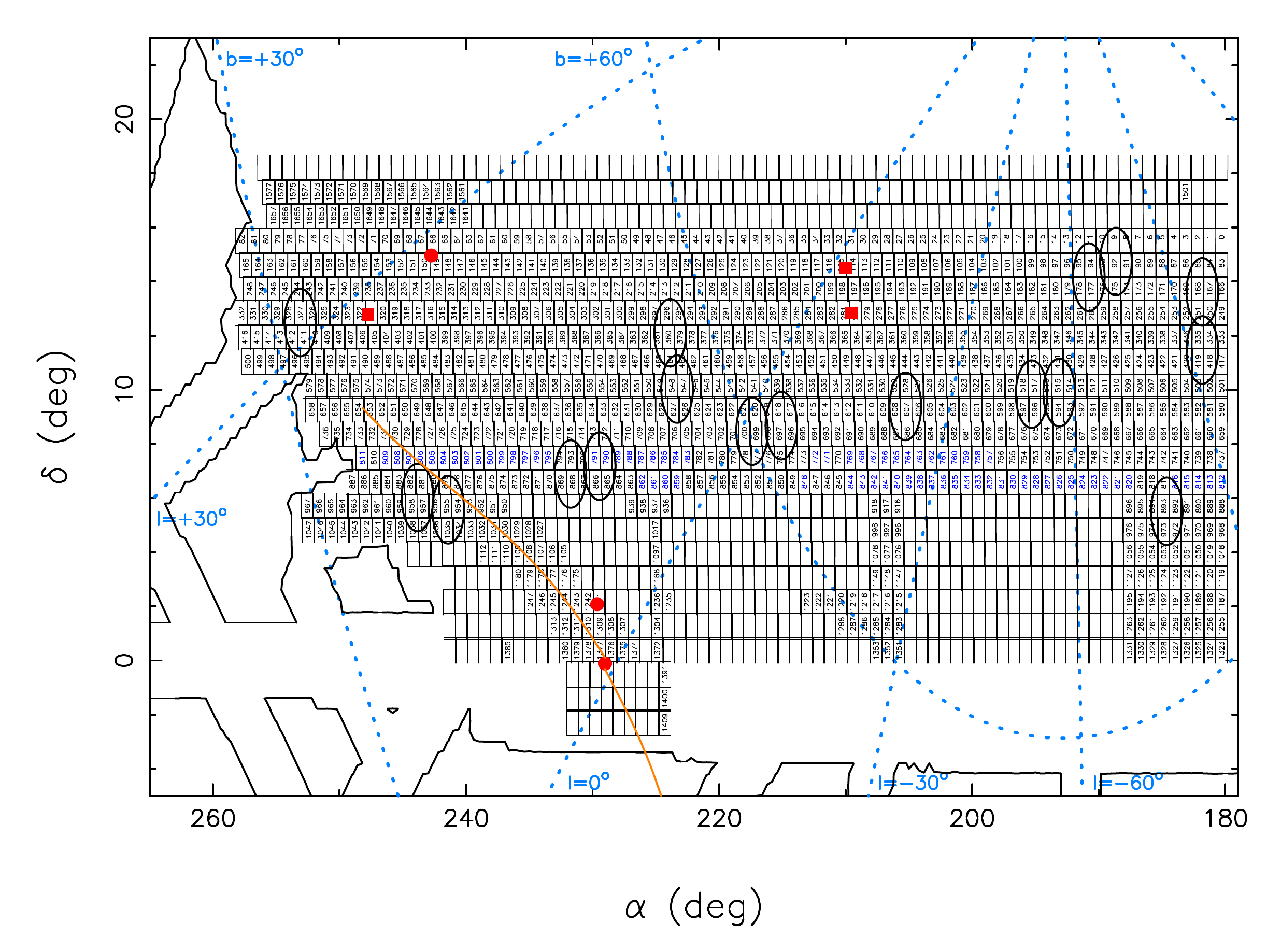}
\caption{Coverage of the \emph{Pristine} survey as of September 2016. The coverage is shown here in equatorial coordinates, but a Galactic coordinate system is overplotted as dotted blue lines. Each rectangle represents a 1 deg$^2$ MegaCam field. Fields with numbers have been observed whereas as those shown without numbers are planned for future semesters. Fields observed with our generic constraints (a single 100s exposure with IQ$ < 0.8"$) have a field number written in black while those with a blue number were observed for longer under poorer conditions ($2\times100$\,s, IQ$ < 1.5"$). Red symbols represent the location of known Milky Way satellites. The squares correspond to the 3 faint dwarf galaxies Hercules, Bo\"{o}tes~I, and Bo\"{o}tes~II, whereas the dots highlights the locations of Pal~5, Pal~14, and M~5. The orange curves departing from Pal~5 represent its stellar stream, as per \citealt{Ibata16}. The large black ellipses correspond to SEGUE fields that overlap with the current \emph{Pristine} footprint and are used for calibration. Finally, the complex black polygon indicates the edges of the SDSS footprint.}
\label{fig:map}
\end{figure*}

In order to build a sample of several tens of stars with $\FeH < -4.0$, one of the science goals of \emph{Pristine}, we aim for \emph{Pristine} to cover at least 3,000 deg$^2$ of the Galactic halo. Since March 2015, we have accumulated more than 1,000 MegaCam pointings with CFHT. These observations were performed through the CFHT queue and scattered through semesters 2015A and 2016A for $\sim 50$ hours of observing time. The state of the survey is presented in Figure~\ref{fig:map} as of August 2016 (i.e. the end of semester 2016A) and compared to the SDSS footprint from which we get broad-band $g$-, $r$-, and $i$-band photometry. The \emph{Pristine} footprint currently covers about 1,000 deg$^{2}$, stretching from $\alpha = 180\deg$ to $\alpha \sim $250--260$\deg$ (or $b \sim 30^{\circ}$) and from $\delta = +6\deg$ to $+16\deg$, with some currently non-contiguous coverage throughout the $-3\deg < \delta < +18\deg$ range. The footprint was chosen to be accessible from both hemispheres for the spectroscopic follow-up and to include a variety of known stellar halo substructures. It includes a portion of the Sagittarius stream and of the Virgo overdensity \citep[see][for an SDSS map of these regions]{Belokurov06a,Lokhorst16}, along with 3 globular clusters (NGC5904/M5, Pal~5, and Pal~14), most of the Pal~5 stellar stream and 3 faint dwarf galaxies (Bo\"{o}tes~I, Bo\"{o}tes~II, and Hercules). Furthermore, it covers 17 SEGUE spectroscopic fields that we use to calibrate the \emph{CaHK}-to-[Fe/H] relation.

Most fields were observed under good seeing (IQ$ < 0.8"$) and during cloudless, bright\footnote{MegaCam bright nights still correspond to $< $60\% moon illumination since the truly bright time is dedicated to the spectro-polarimeter ESPaDOnS (an Echelle SpectroPolarimetric Device for the Observation of Stars) and the Wide-field InfraRed Camera WIRCam.} nights, even though the conditions were not necessarily photometric. For these fields, a single 100-second integration was performed. The survey depth is driven by the magnitude-limit g$ \sim 21.0$ at signal-to-noise S/N$ \sim 10$ for which it will be possible to easily obtain low- and medium-resolution spectroscopic follow-up, combined with the goal of reaching red giant stars at the virial radius of the Milky Way \citep[$\sim$ 200 - 250 kpc, e.g.,][]{Springel08}. For a subset of the fields (shown with blue numbers in Figure~\ref{fig:map}), a low ranking in the queue led to a change of our strategy. For these fields, we requested less stringent limits on the image quality ($1.2 < \textrm{IQ} < 1.5"$), but we instead performed $2\times100$\,s sub-exposures, ensuring reduced data of similar quality. Fields observed with $\textrm{IQ} > 1.5''$ are removed from the survey and are re-observed.

\subsection{Data reduction}\label{sec:datared}
Images are downloaded directly from the CFHT servers. They are already preprocessed by the Elixir pipeline \citep{Magnier04} and have been de-biased, flat-fielded, and de-trended. The next data reduction steps are performed by the \emph{Pristine} collaboration with the Cambridge Astronomical Survey Unit pipeline \citep[CASU,][]{Irwin01} specifically tailored to CFHT/MegaCam images \citep{ibata14}. First, a confidence map is built from a set of CaHK flat-fields downloaded from the archive. The astrometry of the images is then performed against SDSS stars. The very low density of sources on each CCD for these short and narrow-band observations forces us to use all sources detected in the frame down to S/N$ = 5$. Despite this, most CCDs converge on an astrometric solution that is good at the $0.2"$ level and more than sufficient for \emph{Pristine} science. For the small subset of fields observed twice under poorer conditions, the two exposures of a given field are then stacked. Aperture photometry is performed similarly for all single or stacked images before the resulting catalogues are cross-identified with the SDSS DR12. Multiple observations of the same star are removed from the master catalogue during the cross identification, as well as any stars that are flagged in SDSS $g$, $r$ or $i$ photometry to be saturated, have deblending or interpolation problems, are suspicious detections, or are close to the edge of a frame.

We note that in combining both datasets we inherit the bright limit from the SDSS survey, which is typically around $g_0 \approx 14$. The saturation limit for the \emph{CaHK} photometry itself lies typically around \emph{CaHK}$_0 \approx 12$ for our fields, corresponding to $g_0 \approx 11.5$ (naturally dependent on the colour of the star).

As a final step of the data reduction process, we deredden the data following the \citet{Schlafly11} recalibration of the \citet{Schlegel98} maps. E. Schlafly (private communication) kindly determined the extinction coefficient of the CaHK filter, $A_{CaHK}/E(B-V)_\mathrm{SFD} = 3.924$, based on the filter curve.

\begin{figure}
\includegraphics[width=\linewidth]{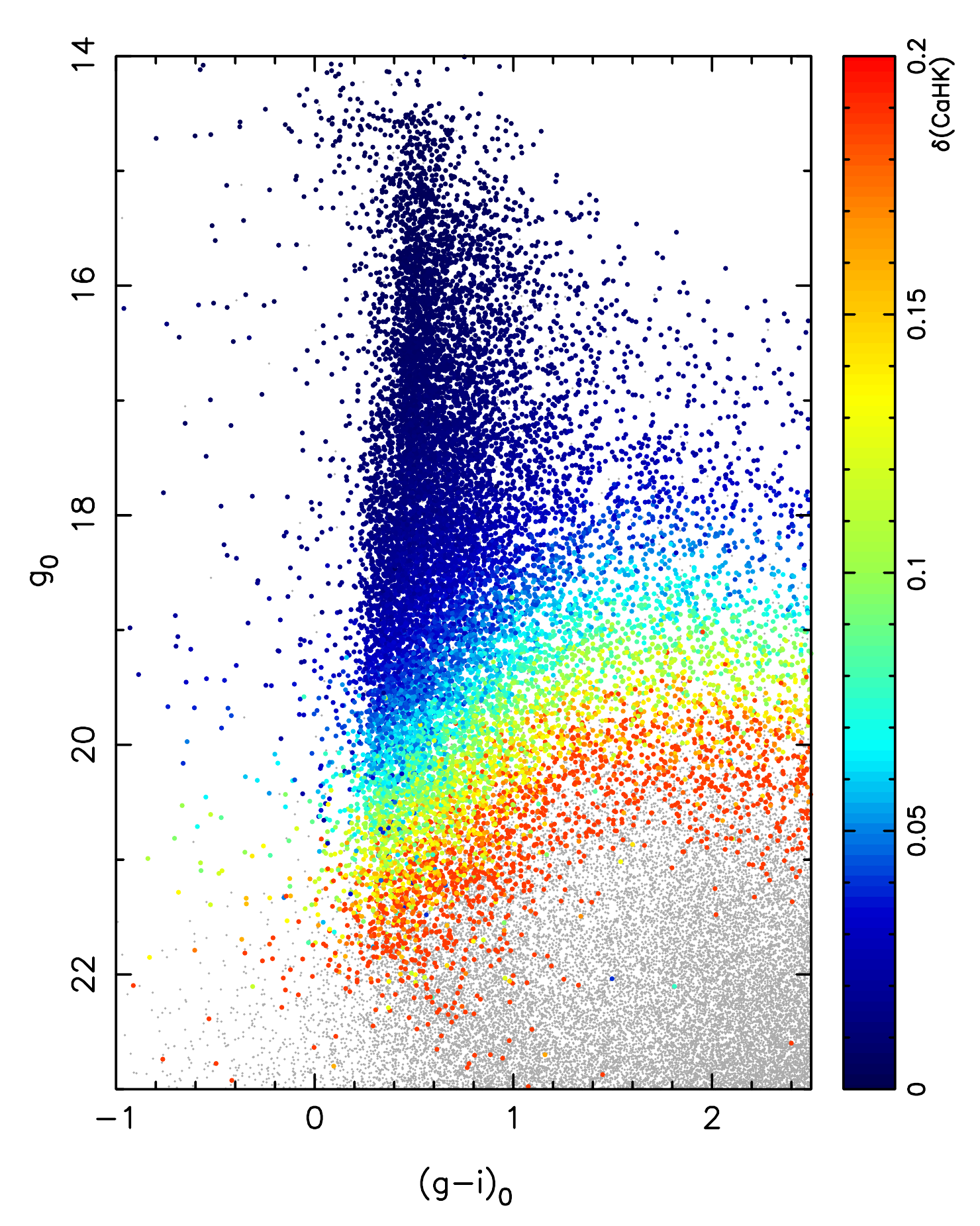}
\caption{SDSS CMD of a $\sim 6$-deg$^2$ region around $(\alpha,\delta) = (236.5\deg,+11.0\deg)$. Extinction variations are minimal in this halo region. SDSS stars are colour-coded by the uncertainties in the \emph{Pristine} CaHK photometry. In this region, typical of the whole survey, the data reach S/N$ \sim 10$ at $g_0 \approx 21.0$ for blue stars close to the turn-off, as originally planned. It corresponds to shallower depths for redder stars, with $g_0 \sim 20.0$ for the same S/N at $(g-i)_0 \sim 1.5$.}
\label{fig:uncertainties}
\end{figure}

The typical photometric uncertainties reached by the \emph{Pristine} CaHK observations are illustrated in Figure~\ref{fig:uncertainties} for a randomly chosen patch of $\sim 6$ deg$^2$ around $(\alpha,\delta) = (236.5\deg,+11.0\deg)$. Given the blue wavelength of the CaHK filter, it is neither a surprise that the magnitude at which a given S/N is reached varies strongly as a function of colour, nor that the blue stars have a higher S/N at fixed magnitude. We reach our intended goal with, overall, S/N$ = 10$ at $g_0 \sim 21.0$ for a main sequence turnoff star and $g_0 \sim 20.0$ for a tip of the giant branch halo star, which corresponds to a distance of $\sim 250$\,kpc for the latter.

 \begin{figure*}
\includegraphics[width=\linewidth]{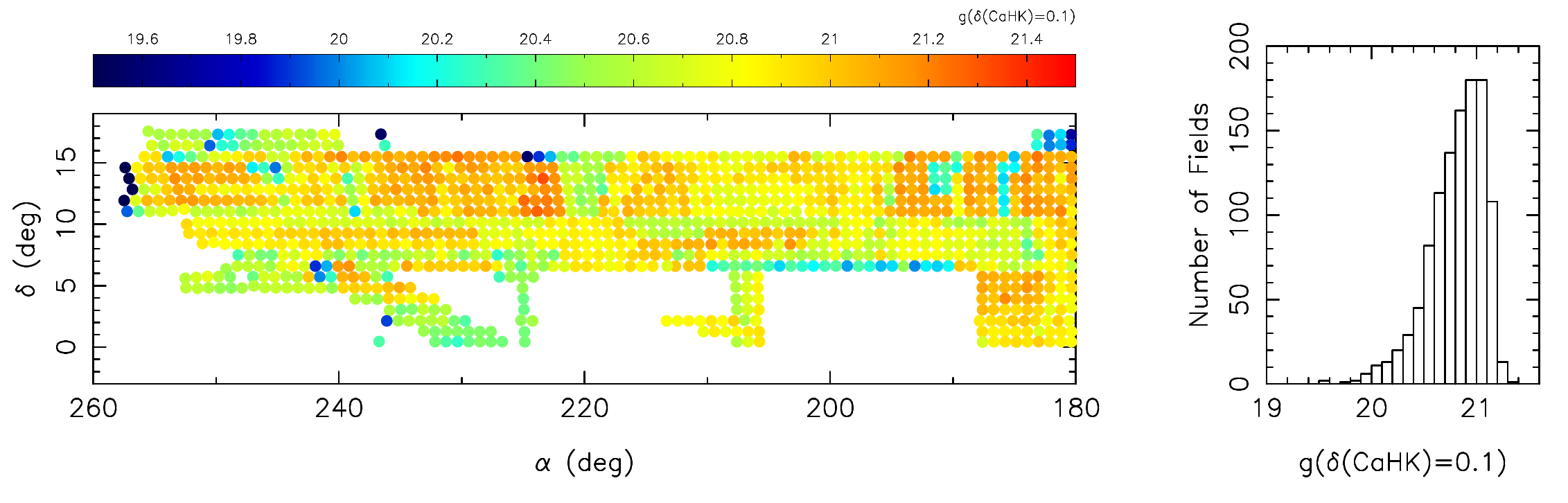}
\caption{Left: Median $g$-magnitude depth reached in each field by stars with $0.2 < (g-i)_0 < 0.6$ and S/N$ \approx 10$ in the CaHK calibrated observations. Right: The distribution of depth values. The median of the distribution is $g_0 = 20.9$ and most of the survey is fairly homogeneous with S/N$ = 10$ depths in the range $20.5 < g_0 < 21.2$. }
\label{fig:depths}
\end{figure*}

Figure~\ref{fig:depths} shows the $g$-band magnitudes reached for CaHK S/N$ = 10$ in the $0.2 < (g-i)_0 < 0.6$ colour band for all fields observed until now. These depths are determined after the calibration is performed (see below). Depth variations are inevitable for a survey like \emph{Pristine} but the median depth is $g_0 = 20.9$ and, although there is a handful of fields in a tail of shallower depths, most of the observed fields have S/N$ = 10$ in the range $20.5 < g_0 < 21.2$. We may revisit some of the significantly shallower fields in the future to bring them closer to the bulk of the survey.

\subsection{Calibration}\label{sec:calib}
We enforce a two-step process to perform a relative calibration of the $\sim 1000$ \emph{Pristine} fields. First we determine a zero point for every field, then we apply a ``flat-field'' at the catalogue level to account for subtle variations of the \emph{CaHK} magnitudes as a function of the position on the field of view.

\subsubsection{Field-to-field calibration}\label{sec:calib_ftf}
\begin{figure}
\begin{center}
\includegraphics[width=1\linewidth]{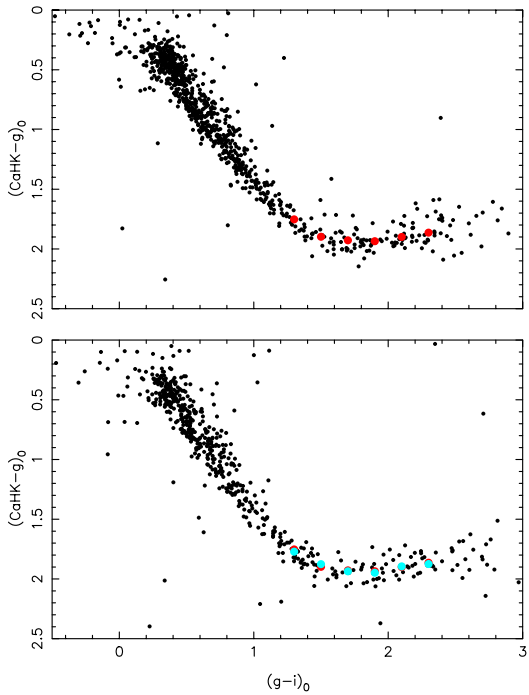}
%using code /Volumes/Data/Pristine_2016A_temp/calibration_16A_redDwarf.e to generate that plot.
\caption{(\emph{CaHK}$-g)_0$ vs. $(g-i)_0$ colour-colour plot for the chosen reference MegaCam field (top) and a random survey field after the field-to-field calibration is performed (bottom). The stellar locus is made of metal-rich disk stars and stars of lower metallicity will be spread above it. To determine the zero point of each field, we use the shape of the stellar locus for foreground red dwarf stars in the range $1.2 < (g-i)_0 < 2.4$ and minimise the $\chi^2$ between the median location of the stellar locus in a reference field (the red points) and those of the point of interest after applying the calibration offset (the blue points for the field represented in the bottom panel).\label{fig:globalCalib}}
\end{center}
\end{figure}

There are no good photometric standard stars for the CaHK filter, so we rely entirely on calibrating all the fields relatively to each other. To do so, we select a reference field. Figure~\ref{fig:globalCalib} presents the dereddened $\left(\left(g-i\right)_0,\left(\emph{CaHK}-g\right)_0\right)$ colour-colour space for our chosen reference MegaCam field (field 251, top panel) and a random survey field (field 23, bottom panel). In these panels, built from stars with \emph{CaHK} uncertainties lower than 0.1, the sequence that extends from (0.3,0.3) to (2.4,1.7) is traced by the bulk of stars and as such defines the stellar locus. This stellar locus is naturally populated by the most abundant types of stars, main-sequence turnoff stars in the blue and mainly nearby, roughly solar-metallicity, disk red dwarf stars in the red. Stars above the stellar locus will have decreasing metallicities, as will be shown later in Section 3. Outliers can be variable stars, stars with \emph{CaHK} in emission or, to the blue of the bluest edge of the sequence, blue horizontal or blue straggler stars.

For each field, we determine the median $(\emph{CaHK}-g)_0$ colour for every 0.2 bin in $(g-i)_0$ for the range $1.2 < (g-i)_0 < 2.4$\footnote{The colour range over which the average location of the stellar locus is determined is particularly important as the colour of the stellar locus needs to be invariant with the location on the sky in the chosen range to be a good calibrator (see Section \ref{sec:photmet_radec}).}. Since the SDSS $g$ and $i$ bands are already calibrated, the median $(\emph{CaHK}-g)_0$ colour of each $g-i$ bin will only vary from field to field because of an unknown zero point offset in \emph{CaHK}. For a given field, we therefore determine the \emph{CaHK} offset necessary to minimise the $\chi^2$ between the median colour points of the locus of this field and that of the reference field, as illustrated in the top panel of Figure~\ref{fig:globalCalib}. \emph{CaHK} magnitudes of this field are then corrected by this most likely offset to put them all on the same zero-point. An example of the result after correction is shown in the bottom panel of Figure~\ref{fig:globalCalib} for field~23.

\begin{figure*}
\begin{center}
\includegraphics[width=1\linewidth]{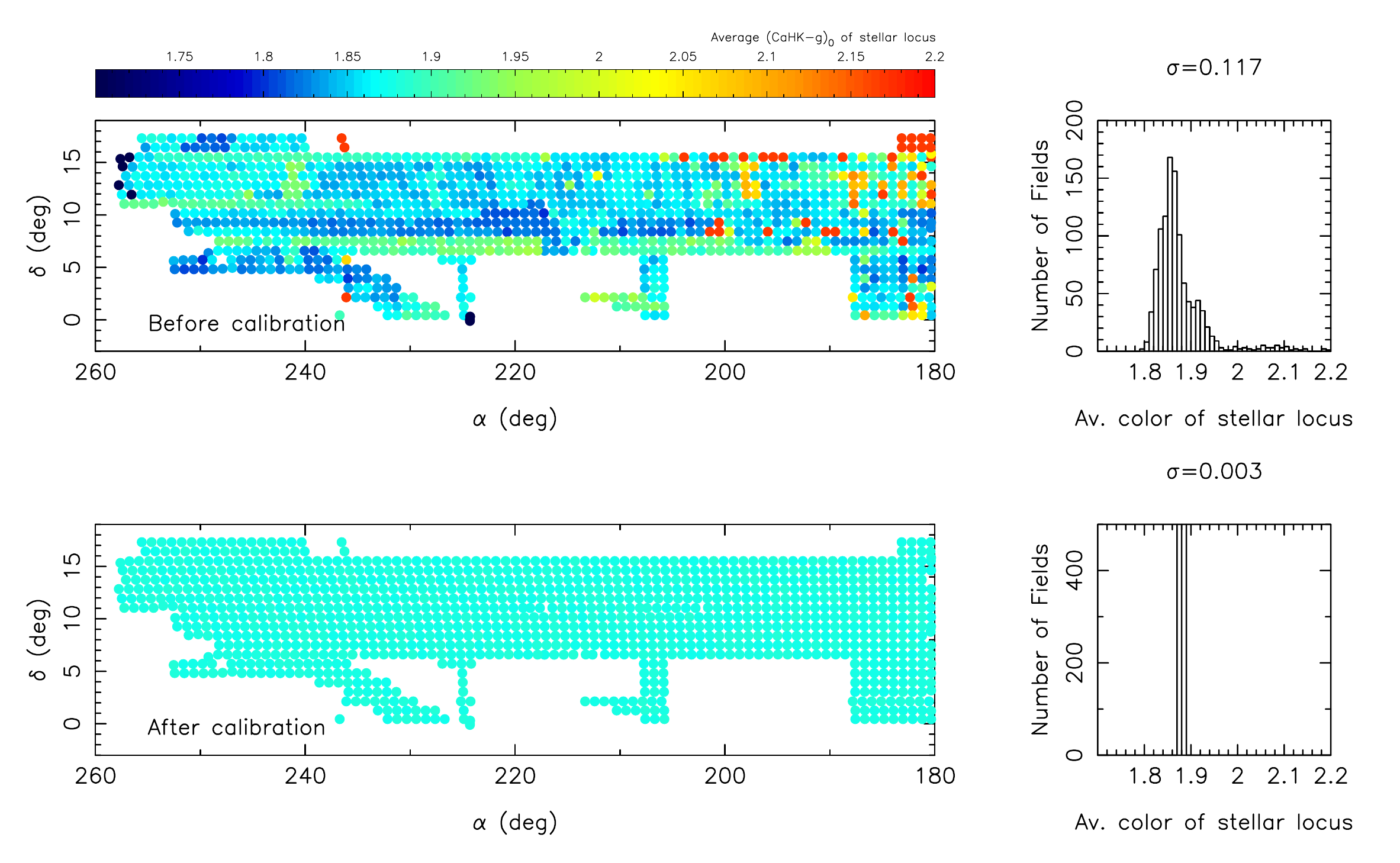}
%using code /Volumes/Data/Pristine_2016A_temp/calibration_16A_redDwarf.e to generate that plot.
\caption{\label{fig:mapCalib}Left: Maps of the average colour of the stellar locus in the colour range $1.2 < (g-i)_0 < 2.4$ before and after the zero point calibration (top and bottom, respectively). Right: The distribution of the average colour of the stellar locus before and after the calibration (top and bottom, respectively). Note the much narrower distribution after the calibration. The respective dispersions of the distributions around their means are indicated above the panels.}
\end{center}
\end{figure*}

To illustrate the quality of this internal calibration, Figure~\ref{fig:mapCalib} shows the average colour of the stellar locus in the $1.2 < (g-i)_0 < 2.4$ (effectively, the average of the six median points determined for a given field) before and after the calibration. It is clear that the calibrated map is much flatter than the uncalibrated one, as it should be by construction. The dispersion around the mean of the distribution drops from 0.137 magnitudes down to only 0.003 magnitudes. Of course, this dispersion does not account for \emph{external} sources of uncertainties on the calibration but confirms that our routine to calibrate the data relatively leads to very good \emph{internal} consistency.

\subsubsection{``Flat-field'' calibration}\label{sec:calib_ff}
The broad-band MegaCam images pre-processed by Elixir are known to suffer from magnitude offsets that depend on the location in the field of view \citep[e.g.,][]{ibata14}. This is also the case for our CaHK images\footnote{For this filter, it could also be the consequence of the throughput and shape of the filter curve varying slightly with the location on the MegaCam field of view. We further note that the dataset was ensured to be homogenised after an early 2016 update to Elixir, as all the \emph{Pristine} images were kindly reprocessed by CFHT staff with the updated version of the Elixir pipeline.}.

\begin{figure*}
\begin{center}
\includegraphics[width=1\linewidth]{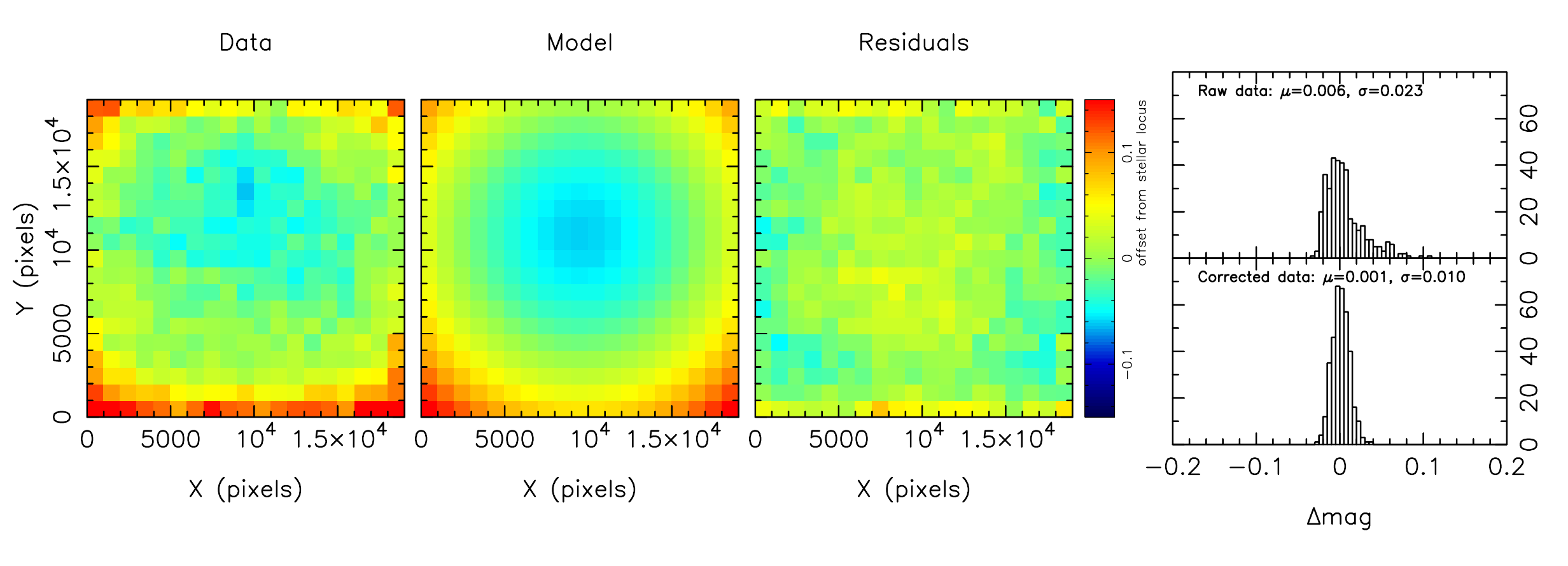}
%using code /Volumes/Data/Pristine_2016A_temp/FoV_calibration_16A.e to generate that plot.
\caption{\label{fig:FFCalib}Left: Median \emph{CaHK} offset of a star from the stellar locus as a function of its location on the MegaCam field of view. Middle-left: Offset model fitted to the data. Middle-right: Residuals after subtraction of the model from the data. Right: Distribution of offset values before the ``flat-field'' calibration (top; mean of 0.006 and a dispersion of 0.023 magnitudes), and that of offset values after correction by the model (bottom; mean of 0.001 and a dispersion of 0.010 magnitudes).}
\end{center}
\end{figure*}

To measure and correct for the remaining variations after the data has gone through the Elixir pipeline, we examine the location of all survey stars with small uncertainties ($\delta(\emph{CaHK}) < 0.05$) in the colour-colour plane of Figure \ref{fig:globalCalib}. We determine the median $(\emph{CaHK}-g)_0$ colour of the stellar locus for bins of 0.01 in $(g-i)_0$ for the $1.2 < (g-i)_0 < 2.4$ range, similarly to what was done in sub-section~\ref{sec:calib_ftf} but on a much finer scale. Splining these median points leads to a model of the median position of the stellar locus. The $(\emph{CaHK}-g)_0$ offset of any of the considered stars from the splined stellar locus is then calculated and shown in Figure~\ref{fig:FFCalib} binned as a function of that star's location on the MegaCam field of view. This binned map reveals small offsets that smoothly vary over the physical location of a star on the MegaCam image. Since the SDSS $g$-band magnitudes should not depend on its position on the \emph{Pristine} field of view, it is safe to assume that these variations are due to an improper calibration of the CaHK images. The top-right histogram of Figure~\ref{fig:FFCalib} shows that the mean offset is close to 0.0 but that it has a tail of high values that stems from the pixels at the edges of the field of view. It is therefore important to correct for this effect to ensure as smooth a survey as possible.

We fit to the data a quadratic model defined by

\begin{eqnarray}
CaHK_\mathrm{off}(X,Y|X_0,Y_0,A,B) = A R^2+B,\\
\textrm{with } R = \sqrt{\left(\frac{X-X_0}{19000}\right)^2+\left(\frac{Y-Y_0}{19000}\right)^2}.
\end{eqnarray}

\noindent Here, \emph{CaHK}$_\mathrm{off}$ is the measured offset from the median colour of the stellar locus, $X$ and $Y$ are the MegaCam global pixel coordinates, $X_0$ and $Y_0$ define the center of the model in those coordinates, $A$ is the amplitude of the correction and $B$ an offset. $\chi^2$ fitting yields the following best parameters: $(X_0,Y_0) = (9600,11000)$, $A = 0.2$, and $B = -0.03$. Subtracting this model from the data flattens the data offsets as can be seen in the third panel of Figure~\ref{fig:FFCalib}. There is still some low level structure in the map but these are much weaker than they were before the correction. This is also made evident by the distribution of pixel values before and after calibration, whose mean and standard deviations change from $(\mu,\sigma) = (0.006,0.023)$ to $(0.001,0.010)$ magnitudes.

\subsection{Global estimation of remaining systematic uncertainties}\label{sec:sysunc}

\begin{figure}
\begin{center}
\includegraphics[width=0.8\linewidth]{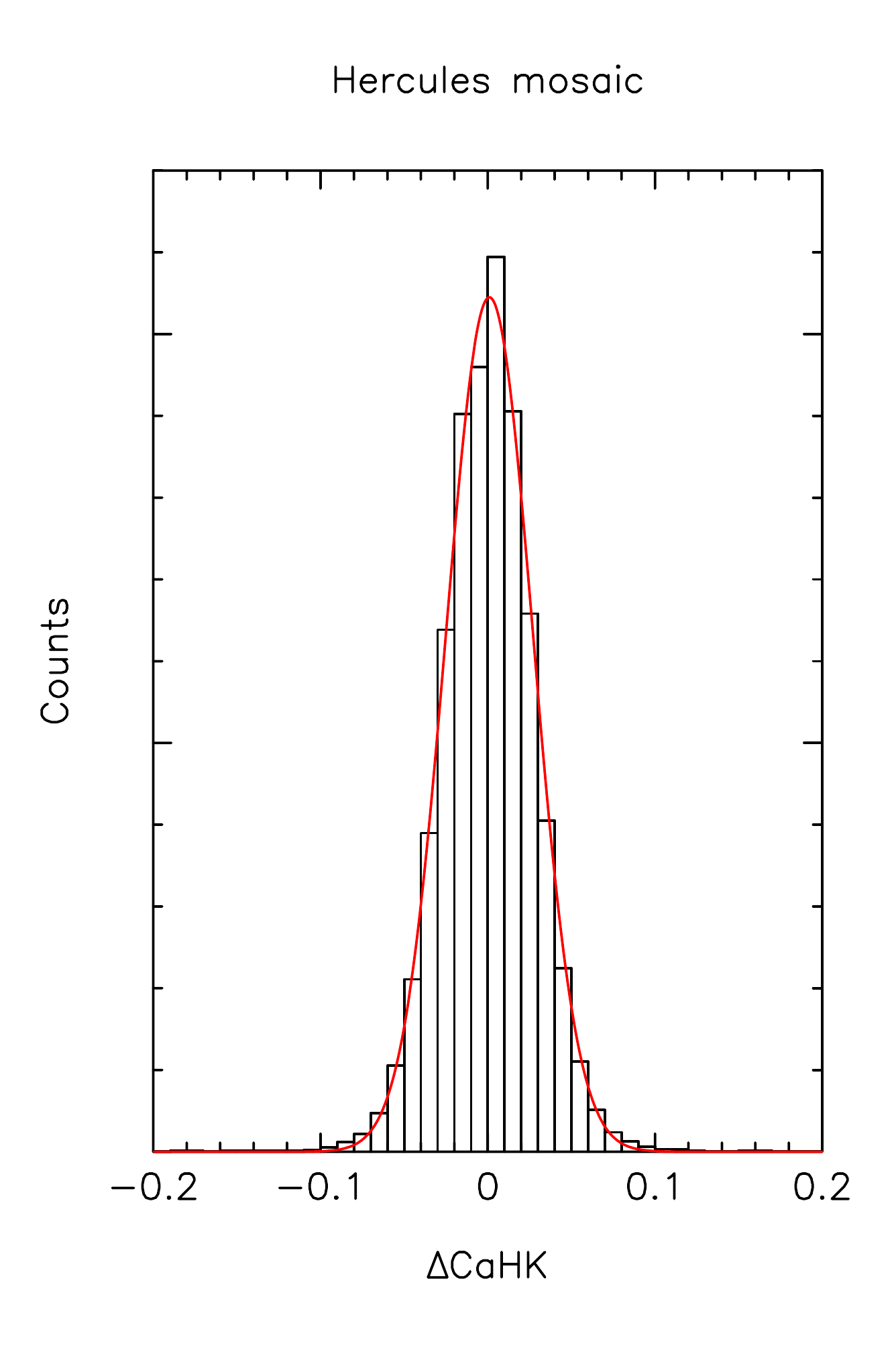}
\caption{\label{fig:diff_mag}Using the mosaic of overlapping Hercules fields to study the distribution of magnitude differences between two \emph{CaHK} measurements of the same star. Shown here is the Gaussian fit to the distribution in red, which is centred at 0.000 and has a width of 0.026, implying an uncertainty floor of 0.018 magnitudes.}
\end{center}
\end{figure}

In order to assess the quality of the \emph{Pristine} CaHK calibration and ascertain the scale of any remaining systematics, we consider a mosaic of fields purposefully designed around the Hercules dwarf galaxy. This mosaic of 25 fields (20 of which were observed) are designed to be shifted positively and negatively by a third and two thirds of a field in both the RA and the DEC directions from a central field centred on Hercules. All the mosaic fields were observed the same night and the comparison of the multiple measurements of a star yields the histogram in Fig~\ref{fig:diff_mag}. In this case, we use all stars with \emph{CaHK} uncertainties less than 0.02 to infer a dispersion of 0.026 magnitudes, which implies an uncertainty floor of 0.018 by dividing this number by $\sqrt{2}$. Since this represents an ideal case, with fields observed during a single night and very similar conditions, we also check the small overlapping region between fields observed on different semesters and similarly conclude that the \emph{CaHK} uncertainty floor is 0.02 magnitudes. We add this number in quadrature to the original uncertainty measurements.

\section{From \emph{CaHK} magnitudes to a photometric metallicity scale}\label{sec:segue} 

Over 17,500 stars in our footprint also have SDSS/SEGUE spectra along with radial velocities and metallicities derived from these, which we use to calibrate the \emph{Pristine} metallicity scale. For our calibration and verification purposes we make some further selection to ensure we are only using the most robust of these data. We select stars from either the SDSS/legacy, SDSS/SEGUE1, or SDSS/SEGUE2 fields that have an average signal-to-noise ratio per pixel larger than 25 over the wavelength 400--800 nm, a derivation of log(g), a radial velocity uncertainty $< 10\kms$, an adopted [Fe/H] uncertainty lower than 0.2, an adopted T$_{\rm eff} < 7000$ K and only nominal `n' flags (with an exception for stars that show the `g' or `G' flag indicating a mild or strong G-band feature). We additionally require that the \emph{Pristine} detection of the star indicates it is indeed a point source (CASU flag of $-1$ in our photometry) and that the uncertainty in the \emph{CaHK} magnitude is $< 0.05$. To mitigate the sparsity in very cool low-metallicity stars, we complement the SDSS sample with the red giant stars of the Bo\"{o}tes I dwarf galaxy, as studied by \citet{Feltzing09,Gilmore13,Ishigaki14} and \citet{Frebel16}. In total, this leaves us with 7673 stars to calibrate our survey with.

In Appendix A. we compare how the various [Fe/H] estimates from the SDSS stellar parameter pipeline (SSPP; see \citealt{Lee08} for an overview description) behave in our \emph{Pristine} colour-colour space with the highly temperature-sensitive SDSS $(g-i)_0$ colours and a combination of these and the \emph{Pristine} \emph{CaHK} magnitude on the $y$ axis (the same combination as introduced in Figure \ref{fig:synt}). This can be seen as an independent testing of the different pipelines. 

It must be noted that in all cases the comparison is based on [Fe/H] values, whereas naturally our filter is mostly sensitive to Ca lines. For most SDSS methods compared, the [Fe/H] value given is however also influenced by features in the stellar spectrum that follow elements besides Fe. Also, we remind the reader that the comparison with synthetic spectra, such as shown in Figure \ref{fig:synt}, is based on assumptions on the [Ca/Fe] value with [Fe/H]. Specifically, we use a scale for [Ca/Fe] that linearly rises from 0.0 at [Fe/H] = 0.0 until $+0.4$ at [Fe/H] = $-1$ and then stays at this level for lower metallicities. We note that, for any stars with a discrepant [Ca/Fe], the calibration might be less accurate.

Although the standard adopted [Fe/H] from the SSPP -- called FEHADOP and a combination of all methods -- performs well for the bulk of the stars, we find that for the lowest metallicity regime a method called FEHANNRR provides a better result. The FEHANNRR metallicity estimate is neural network (NN) based using an initial Principal Component Analysis compression of the data and it utilises the full wavelength range of the SDSS/SEGUE spectra. It is further both trained and tested on observed ('real', hence RR) SDSS spectra instead of synthetic spectra \citep{ReFiorentin07}. As shown in Appendix A., with this method the stars with spectroscopic [Fe/H] values close to [Fe/H]$ = -3$ fall close to the synthetic $-3$ line and additionally fewer stars in other parts of the parameter space receive such a low spectroscopic metallicity. Because of its superior behaviour at the lowest metallicity regime, we will adopt this as our standard SSPP spectroscopic metallicity value for the remainder of the paper.

\begin{figure}
\includegraphics[width=\linewidth]{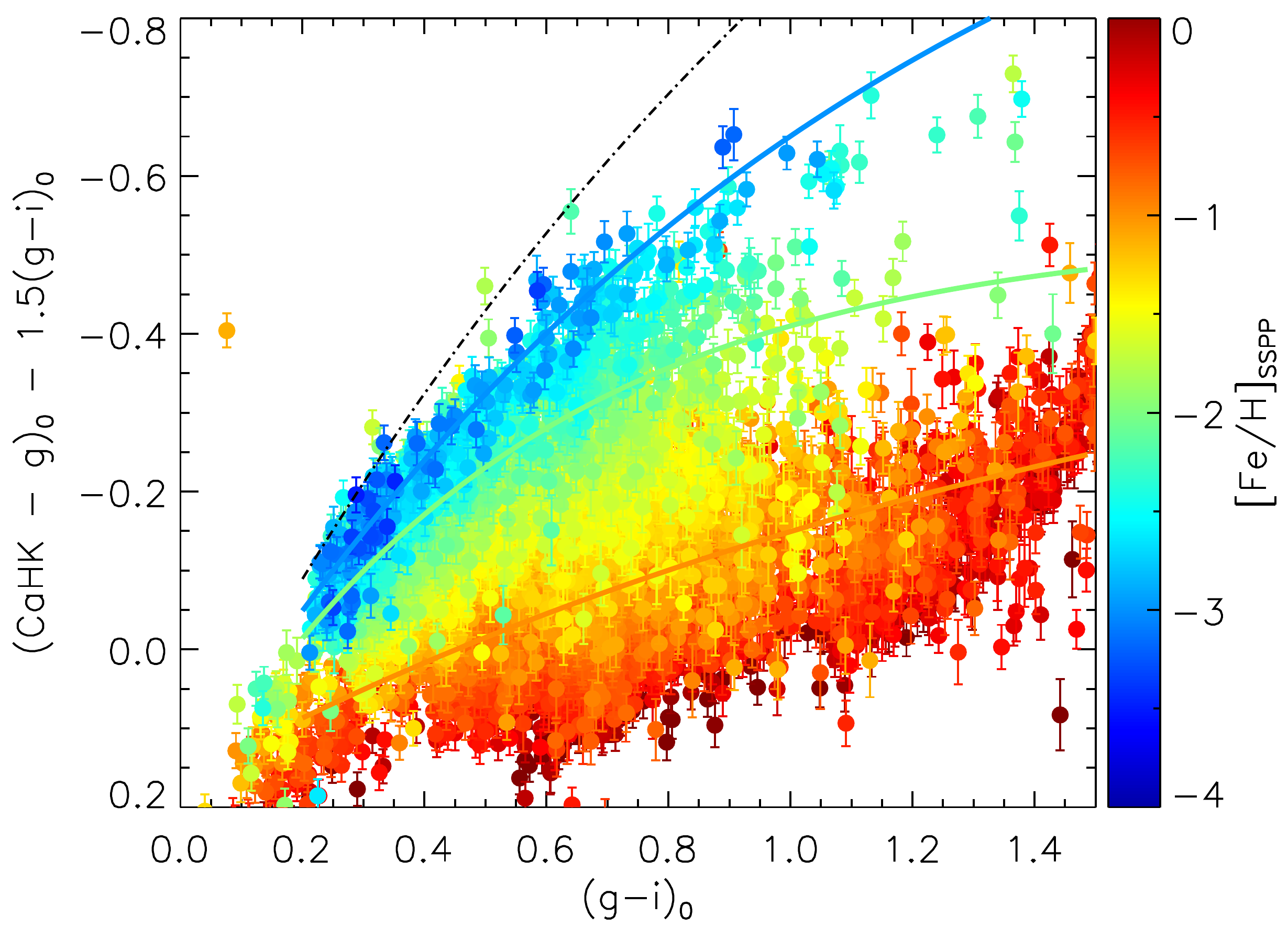}
\caption{The sample of stars within the \emph{Pristine} footprint overlapping with SDSS/SEGUE spectra, plotted in the SDSS-\emph{Pristine} colour-colour space as introduced in Figure \ref{fig:synt}. As described in the text, several cuts are applied to this sample to remove contaminants and uncertain measurements. The stars are colour-coded by their [Fe/H] from the SDSS stellar parameter pipeline, SSPP. The coloured lines are using the same colour-scale and represent the fits to synthetic spectra model predictions for stars with $\FeH = -1.0$, $-2.0$, $-3.0$, and with no metal lines as presented in Figure \ref{fig:synt} (orange, green, blue, and black, respectively); the latter is offsetted by 0.05 to provide an upper limit.}
\label{fig:Segueclean}
\end{figure}

\subsection{Cleaning the contamination}\label{sec:cont}

Since the main \emph{Pristine} goal is to isolate particularly under-represented objects in the survey, we must remove as many contaminants as possible. We aim to weed out two particular sources of contamination: 
\begin{itemize}
\item{\emph{Variable stars \& quasars}: Because the \emph{Pristine} narrow-band photometry and the SDSS broad-band photometry are not taken at the same time (but instead several years apart), variable sources can significantly scatter around up and down in the colour-colour diagram. To remove these interlopers, we cross-identify our catalogue with the catalogue of $\chi^2$ variability of all Pan-STARRS1 \citep[Panoramic Survey Telescope and Rapid Response System][]{PS1} sources by \citet[][see their equation 1.]{Hernitschek16} and remove every object that has a $\hat{\chi}^2$ value larger than 0.5. This affects mostly the higher temperature stars ($(g-i)_0 < $0.3), but also flags some cooler variable objects.}
\item{\emph{White dwarfs}: Nearby cool white dwarfs can overlap with the temperature and magnitude range of interest as well. As white dwarfs typically show no Ca absorption features, they are a potential source of contamination in our metal-poor samples. However, following \citet{Lokhorst16}, the sparse white dwarf population can be removed effectively from the main population of main-sequence and red-giant stars by removing any star from our sample that has SDSS $(u-g)_0 < 0.6$.}
\end{itemize}

By making these cuts we remove 7\% of our SDSS/SEGUE overlap sample and a similar percentage in our total sample, because they are suspected contaminants. One remaining possible source of contamination amongst our low-metallicity targets are chromospherically active KM dwarfs showing Ca H \& K in emission. Strong chromospheric activity is however observed only among stars younger than 0.1 Gyr \citep{Henry96}. In the HES survey, which was not affected by variability issues, it has been documented that only 43 out of 1771 metal-poor candidates had to be rejected after low-resolution follow-up \citep{Schorck09}. Thus, we conclude that those with features so strong that they (exactly) fill in their absorption lines will comprise a very small fraction of our targets. 

Figure \ref{fig:Segueclean} illustrates the overlapping and cleaned SDSS/SEGUE sample in a colour-colour plot. The coloured lines that are overplotted on the data points are the same as in Figure \ref{fig:synt} and represent the exponential fits to expectations from the integration of synthetic spectra. The line fitting the synthetic spectra with no metals (shown as a dashed black line) is offset by 0.05 upwards, in order to accommodate the typical photometric uncertainty in our data set and act as an upper limit of where we expect our stars to fall. 
 
\subsection{Deriving photometric metallicities}\label{sec:photmet} 

Using the cleaned sample of \emph{Pristine} stars with SDSS spectra in conjunction with our synthetic spectra, we define a metallicity scale to assign a \emph{Pristine} photometric metallicity star to every \emph{Pristine} star. The $(g-i)_0$ and [$\emph{CaHK}-g-1.5(g-i)]_0$ space shown in Figure \ref{fig:Segueclean} is pixelized in square pixels of 0.025 in size and each pixel is assigned the average FEHANNRR metallicity value of all common SDSS/SEGUE--\emph{Pristine} stars. Because of the larger number of stars at higher metallicities, we expect more contamination of higher metallicity stars in lower metallicity bins than the converse. To avoid them dominating the averaging process, we clip outliers with metallicity values 2--$\sigma$ above the average. Pixels without stars are assigned the metallicity value from the closest populated pixel. Additionally, we enforce a monotonic relation in every $(g-i)_0$ column such that, for $\FeH = -1.5$ pixels and above, a pixel cannot have a value exceeding its downwards neighbour. In other words, at lower metallicities we force the behaviour of the grid such that for a fixed temperature (approximated by $(g-i)_0$), a larger \emph{CaHK} flux means a lower metallicity (as expected from Figure \ref{fig:CaHK}). For pixels that are close to the no-metals line, the sampling is very poor. We therefore assign a metallicity $\FeH = -4.0$ to any empty pixel up to 0.2 above that line. We remind the reader that the no-metals line was calculated using synthetic spectra from metallicity \FeH$ = -4$ stellar atmospheres and no heavy metal absorption lines. The extra 0.2 margin is therefore set to make sure that we do not miss the most interesting metal deficient stars because of systematic effects in either of these steps. To give this region of the colour-colour plot \FeH$ = -4$, ensures the calibration of the low metallicity end of the distribution. As a final step, the grid is slightly smoothed by a 2D Gaussian with a 2-pixel standard deviation to lower the noise.

Subsequently, we go through the exact same procedure for a grid where we use $(g-r)_0$ instead of $(g-i)_0$ (not shown), a colour also known to be very temperature sensitive \citep{Ivezic08}, although we find that, compared to an ordering in $(g-i)_0$, it provides slightly less metallicity discrimination power at the lowest temperatures. We therefore adopt the metallicity based on $(g-i)_0$ colour as the standard photometric metallicity in the remainder of this work and use the $(g-r)_0$-based metallicity just as an extra check of the quality of the photometry. For 2\% of the total sample with $\delta_{CaHK} < 0.05$, we find that the two metallicity scales do not agree within 0.5 dex. These stars are discarded from the sample in the rest of this work. For the remaining 98\% of the sample, though, the two photometric metallicities are in good agreement. For over 80\% of the sample the agreement is better than 0.2 dex and for over 50\% better than 0.1 dex. 

\subsection{Quality assessment of the \emph{Pristine} metallicities}

\subsubsection{The effect of photometric uncertainties on the derived photometric metallicities}
To assess the photometric metallicity uncertainties due to the uncertainties on the \emph{CaHK} magnitudes alone, we look again at the stars that are observed multiple times in the mosaic of overlapping fields in the region of the Hercules dwarf galaxy, as described in Section \ref{sec:sysunc}. We find that in the full sample, which represents many different stellar parameters, the \emph{CaHK} uncertainty of 0.026 magnitudes within the Hercules sample corresponds to a median uncertainty in metallicity of $\sigma_\FeH = 0.11$. In the metal-poor regime ($\FeH < -1.5$) $\sigma_\FeH$ increases to 0.20, because a variation in \emph{CaHK} corresponds to a larger difference in \FeH\ in this regime.

\begin{figure*}
\includegraphics[width=\linewidth]{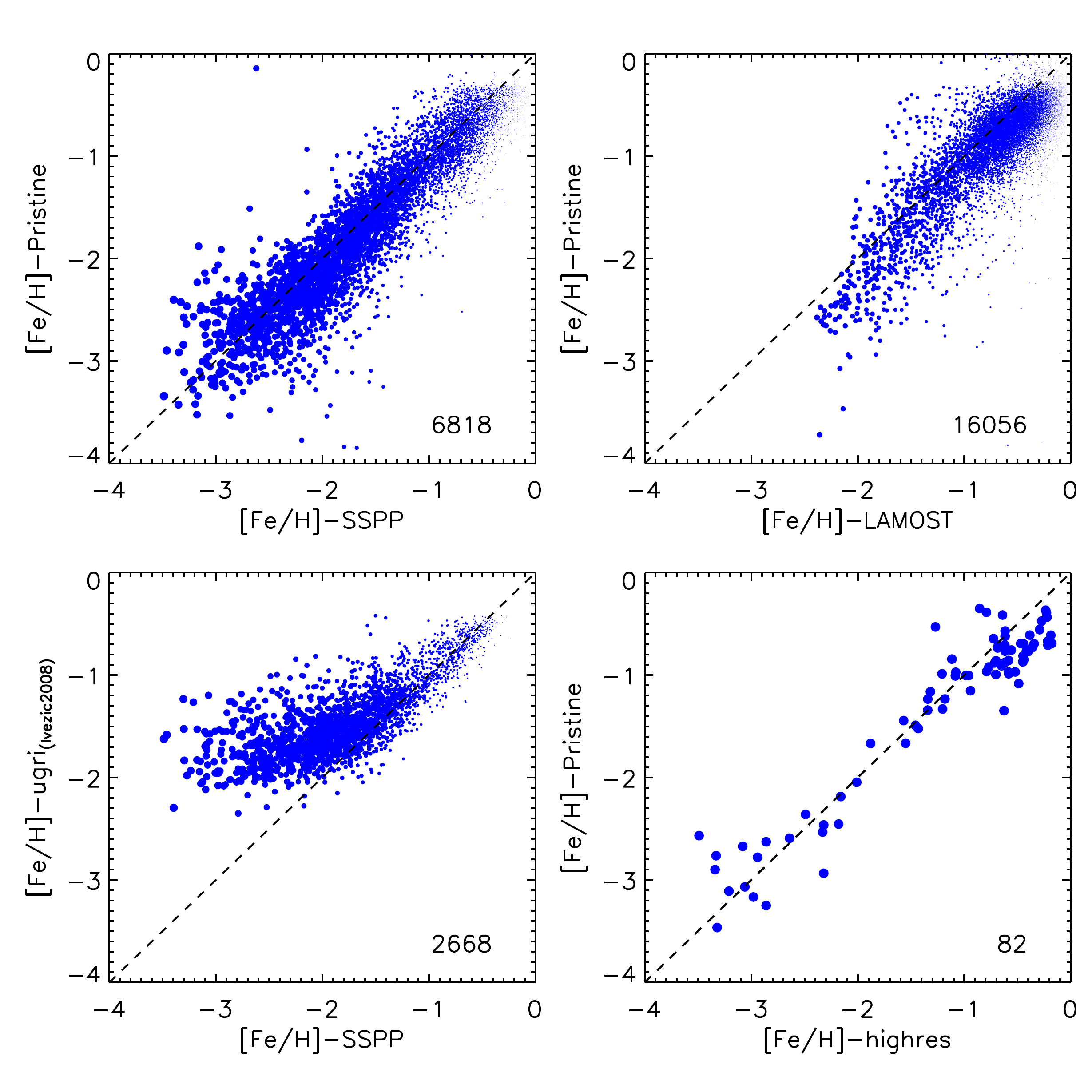}
\caption{Left top panel: \emph{Pristine} photometric [Fe/H] as determined by the photometric calibration described in the text using the \emph{CaHK}$_0$ magnitudes and SDSS $g_0$ and $i_0$, compared to the spectroscopic [Fe/H] as determined from SDSS and SDSS/SEGUE spectra by the FEHANNRR method in SSPP. Right top panel: \emph{Pristine} photometric [Fe/H] compared to the spectroscopic [Fe/H] as determined by the LAMOST pipeline in the LAMOST data release 2 sample. Bottom left panel: Broad-band photometric [Fe/H] as determined from SDSS $u_0$, $g_0$, and $i_0$ alone, following the calibration of \citet{Ivezic08} and their colour cuts for their more stringent sample restricted to stars with $(g-r)_0 < 0.4$ (i.e., the hotter stars close to the turn-off). Because of the extra colour cuts by \citet{Ivezic08}, the bottom left panel has far fewer stars than the top left panel. Bottom right panel: Comparison of \emph{Pristine} photometric [Fe/H] and spectroscopic [Fe/H] from overlapping high-resolution samples. Stars in this sample are taken from SDSS near-infrared high-resolution survey Apache Point Observatory Galactic Evolution Experiment \citep[APOGEE,][]{SDSS16} at the higher metallicities, where we have discarded any star with the APOGEE pipeline flag unequal to zero. At lower metallicities we have cross-correlated the high-resolution datasets of \citet{Aoki13} and \citet{Cohen13} with the footprint of the \emph{Pristine} survey. Finally, we have added high-resolution observations of stars in the Bo\"{o}tes I dwarf galaxy as compiled by \citet{Romano15} \citep[original measurements by][]{Feltzing09,Gilmore13,Ishigaki14} and supplemented with one additional star from \citet{Frebel16}. The symbol sizes on the first three panels are inversely linearly dependent on the metallicity on the x-axis, to allow both a good view on the sparser metal-poor population and dense metal-rich population. The numbers in the panels indicate the number of stars in the sample. }
\label{fig:photmet}
\end{figure*}

\subsubsection{Comparison to spectroscopic datasets}
The left top panel of Figure \ref{fig:photmet} shows the \emph{Pristine} photometric [Fe/H] versus the spectroscopic [Fe/H] from SDSS/SEGUE for stars in common between the two surveys. Stars are further selected to have $0.2 < (g-i)_0 < 1.5$, corresponding to the full colour range from the main-sequence turn-off to the tip of the red giant branch for a metal-poor stellar population, thereby covering most stages of stellar evolution except for the hottest or coolest stars. We find a tight relation, in particular for $\FeH < -0.5$, with a Gaussian fit standard deviation of only $\approx 0.22$ dex throughout the full metallicity range. The Gaussian fit is systematically offset by $\approx -0.08$ dex, in the sense that the \emph{Pristine} metallicities are generally slightly more metal-poor. The dominance of high-metallicity outliers scattering into the lower metallicity regime, as mentioned above, is likely the driving force behind this small systematic offset and we therefore decide not to correct for it. The standard deviation remains stable at the low-metallicity end and increases only slightly to $\approx 0.23$ for $\FeH < -1.5$ and to $\approx 0.25$ for $\FeH < -2.0$. In particular, for this lower metallicity regime, this means that the total uncertainty is not much larger than the uncertainty expected from systematics in the measurement of \emph{CaHK} magnitudes. Even though this set of data is identical to the training set, and thus they are by definition on the same metallicity scale, the small scatter in the full sample does illustrate the excellent metallicity sensitivity of the CaHK filter. Further, more precise, metallicity measurements and the determination of abundance ratios for several elements can be reached by follow-up spectroscopy.

\begin{figure*}
\includegraphics[width=\linewidth]{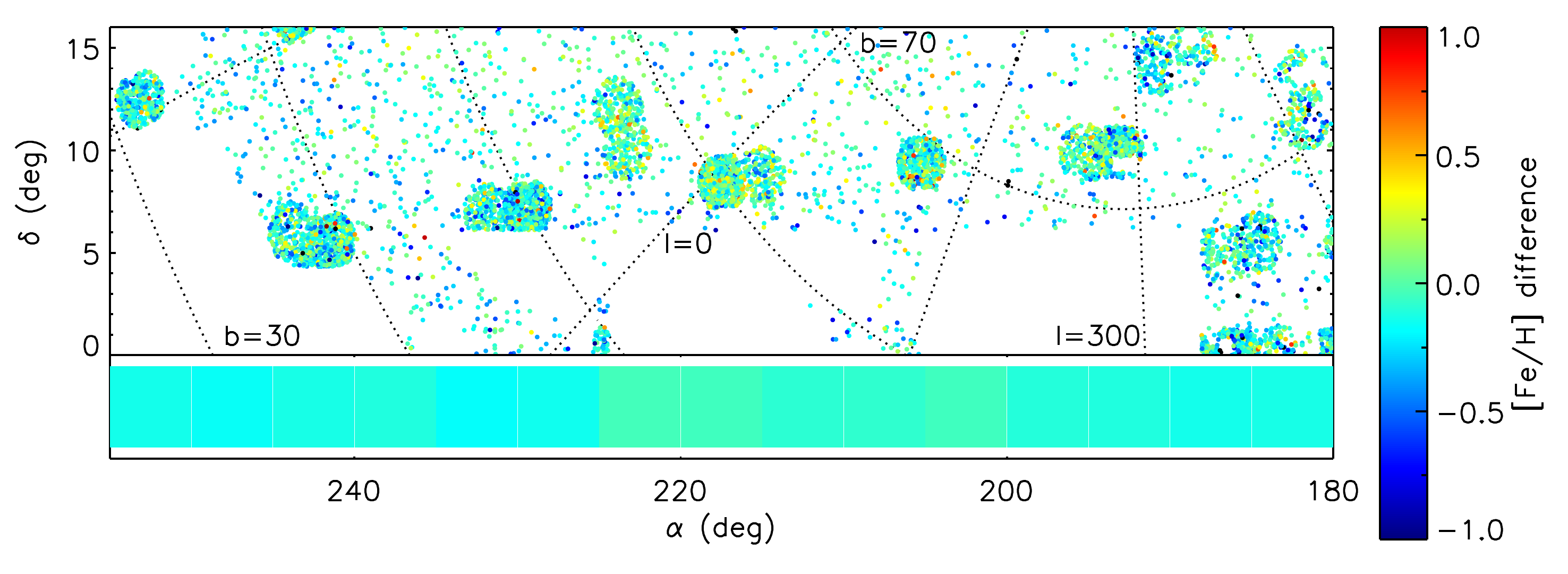}
\caption{Differences between the \emph{Pristine} photometric metallicities and the FEHANNRR values from the SDSS and SDSS/SEGUE stellar parameter pipeline for each star in common between the two surveys. A grid is overplotted with Galactic $l$ and $b$ coordinates as dotted black lines. The lines of constant $l$ are spaced $30\deg$ apart and range from $l = 270$ to $l = 30$ (right to left). The lines of equal $b$ are spaced $10\deg$ apart and range from $b = 30$ to $b = 70$ (bottom left to top right). One can clearly see the locations of the SEGUE fields where the density of overlapping targets is higher. No overall trend is present in the residuals as a function of sky position. The flatness of the overall calibration as a function of Galactic environment is most clearly visible in the bottom panel, where the results are averaged over wide RA range bins.}
\label{fig:photmet_radec}
\end{figure*}

In addition to the SDSS/SEGUE sample, 21,969 \emph{Pristine} stars are also present with measurements for their velocities and metallicities in the DR2 of the LAMOST survey \citep{Cui12, Xiang15, Luo15}. When we apply similar quality cuts to this sample as to our SDSS/SEGUE sample (although with higher tolerance on their derived uncertainties, namely a radial velocity uncertainty $< 70\kms$ and an adopted [Fe/H] uncertainty lower than 0.4 by the LAMOST pipeline) we find a sample of over $ \sim 16,000$ stars that can be used for comparison with the \emph{Pristine} metallicities, of which $\sim 5,700$ with $\FeH < -0.5$. The sample is shown in the top right panel of Figure \ref{fig:photmet}. This comparison leads to very similar conclusions to the comparison with SDSS/SEGUE over the full metallicity range, although it is less well constrained at low metallicities due to the sparsity of low-metallicity stars in the LAMOST dataset. The distribution of metallicity differences between the LAMOST spectroscopic and \emph{Pristine} photometric metallicities has a mean of $-0.12$ and a Gaussian-fit standard deviation of $\approx 0.18$ dex.

We illustrate the clear benefit of additional narrow-band photometry to complement broad-band photometry in the bottom left panel of Figure \ref{fig:photmet}. Here, we have calculated the photometric metallicities from SDSS broad-band photometry \textit{alone}, following \citet{Ivezic08}. Additional colour cuts are applied to the sample, as described by \citet{Ivezic08}. In particular, we only use their most strict colour-cut sample with $0.2 < g-r < 0.4$, which selects mostly hotter stars near the turn-off, for which the calibration performs best. The bottom left panel of the figure therefore contains over $\sim 4,000$ fewer stars than the top left panel even though they start out from the same sample. Despite these extra restrictions, the calibration based on pure broad-band results clearly underperforms over the full metallicity range. Most strikingly, the narrow-band CaHK filter adds a sensitivity to the lowest metallicity regime that cannot be achieved from broad-band photometry.

Finally, we present in the bottom right panel of Figure \ref{fig:photmet} a comparison of \emph{Pristine} metallicities with metallicities derived from high resolution studies. We have cross-correlated the \emph{Pristine} footprint with SDSS-APOGEE \citep{SDSS16}, which mainly yielded stars at [Fe/H]$ > -2$. For this sample, only stars with the APOGEE pipeline flag set to zero are considered. We have sought for stars studied in high-resolution at the low-metallicity end by cross-correlating with the studies of \citet{Aoki13} and \citet{Cohen13}. Additionally, we make use of the coverage of the Bo\"otes dwarf galaxy in our footprint and add high-resolution observations of stars in this galaxy as compiled by \citet{Romano15}, including stars from \citet{Feltzing09}, \citet{Gilmore13}, and \citet{Ishigaki14}, with one additional star from \citet{Frebel16}. The comparison again shows a tight correlation between spectroscopy and \emph{Pristine} photometric [Fe/H] measurements, down to the very low-metallicity regime.

\subsection{The photometric metallicity scale in different Galactic environments}\label{sec:photmet_radec} 

Throughout our footprint the Galactic latitudes observed vary from $b = 30\deg$, close to the disk, to $b = 78\deg$, close to the Galactic North pole. The stellar populations observed therefore change as well; while the low-latitude fields would be completely dominated by more metal-rich stars from the Galactic thin and thick disk, the higher latitude population contains a much larger fraction of metal-poor halo stars. In Figure \ref{fig:photmet_radec} we therefore investigate the robustness of our metallicity scale as a function of the survey footprint and find no systematic trends. The colour range over which the average location of the stellar locus is calculated during the calibration process is particularly important to obtain this result. An initial choice of fitting the stellar locus at $0.4 < (g-i)_0 < 1.2$ led to a systematic bias in the \FeH\ values because of changes in the main-sequence turnoff stellar populations between the halo and the (thick) disk. Physical changes in the colour of the turnoff due to the shifting median metallicity of the stellar population were effectively assigned to zero point offsets, leading to an \FeH\ bias that was a function of the Galactic latitude. The final selection of only foreground dwarf (main sequence) stars in the range $1.2 < (g-i)_0 < 2.4$, whose properties are homogeneous over the sky, resulted in a homogeneous calibration.

\subsection{Purity and completeness of finding low-metallicity stars}\label{sec:rates}

On average over our footprint, we find that in each $\sim $deg$^{2}$ field we have $\sim $7 stars that have $\FeH_\mathrm{Pristine}\leq -2.5$ down to a magnitude of V$ \approx$ 18. When we select stars in common in \emph{Pristine} and SDSS/SEGUE that have a \emph{Pristine} photometric metallicity of $\FeH_\mathrm{Pristine}\leq -2.5$, we find that our success rate at uncovering a star with an SSPP spectroscopic metallicity $\FeH_\mathrm{SSPP}\leq -2.5$ is 51\%. We note that this success rate is based on the SDSS/SEGUE metallicities that are also known to be challenged in the very metal-poor regime \citep[e.g.,][]{Lee08, Aoki13, Aguado16}. Therefore, a more telling statistic is perhaps that 90\% of this subsample has $\FeH_\mathrm{SSPP}\leq-2.0$. 

Additionally, we investigate the completeness of a sample of metal-poor stars selected with \emph{Pristine} photometry and find that 71\% of \emph{all} stars with $\FeH_\mathrm{SSPP}\leq-2.5$ also have $\FeH_\mathrm{Pristine}\leq-2.5$, and 98\% of these have $\FeH_\mathrm{Pristine}\leq-2.0$. 

Similarly, although with smaller sample statistics, we recover 39\% of overlapping SDSS/SEGUE stars with $\FeH_\mathrm{SSPP}\leq -3.0$ by selecting stars with $\FeH_\mathrm{Pristine}\leq-3.0$. Of all stars that have $\FeH_\mathrm{SSPP}\leq -3.0$, we find 78\% to have $\FeH_\mathrm{Pristine}\leq -2.5$. The success rate of uncovering $\FeH_\mathrm{SSPP} \leq-3.0$ among $\FeH_\mathrm{Pristine}\leq-2.5$ selected stars is 8\% and increases to 24\% among $\FeH_\mathrm{Pristine}\leq-3.0$ selected stars. Of the remaining candidates at $\FeH_\mathrm{Pristine}\leq-3.0$, $\sim $85\% are still very metal-poor at $\FeH_\mathrm{SSPP}\leq -2.0$ and can thus not really be considered ``contaminants''. Given that published results from previous or ongoing surveys report a success rate of finding [Fe/H]$ < -3$ stars around 3-4\% \citep{Schorck09,Schlaufman14,Casey15} such a high success rate percentage at 24\% would mean a great success for \emph{Pristine}. 

From these success rates we conclude that a survey of very metal-poor stars based on \emph{Pristine} photometry will be very complete and have a low level of contamination. We can confirm that these rates are indeed achieved from preliminary results of our first spectroscopic follow-up campaign (Youakim et al., submitted).
 
\section{Science with \emph{Pristine}}\label{sec:science}

Below we outline three main science goals of the \emph{Pristine} survey, as well as present some preliminary results from the photometry and the spectroscopic follow-up programs. Detailed results will be presented in upcoming papers.

\subsection{Searching for the most metal-poor stars}

\begin{figure}
\includegraphics[width=\linewidth]{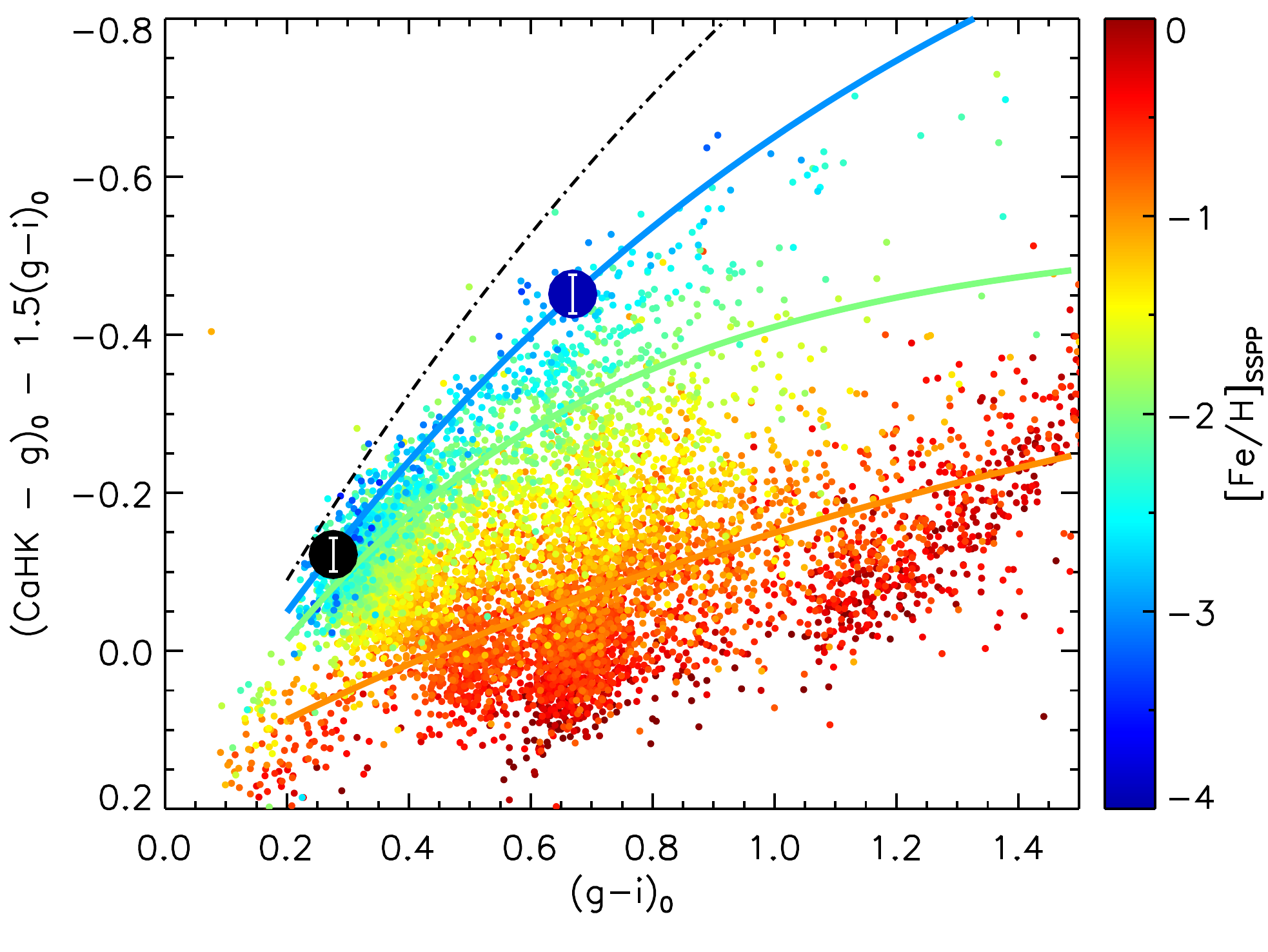}
\caption{From left to right the large coloured filled circles show the locations of SDSS J1742+2531 (black) and SDSS J183455+421328 (darker blue) in the SDSS--\emph{Pristine} colour-colour space as presented before in Figure \ref{fig:Segueclean}. These two stars are known to have $\FeH\leq-4$ and we find that they are located approximately on the $\FeH = -3$ line in the Pristine data. As before, the colour of the symbols indicates their \FeH\ values, determined from higher resolution spectroscopic follow-up by \citet{Caffau13b,Bonifacio15} and \citet{Aguado16}. Smaller dots show the sample of common stars between SDSS/SEGUE and \emph{Pristine}. All these stars are colour-coded according to their SSPP metallicity. The coloured lines represent exponential fits to the symbols of metallicities \FeH$ = -1,-2$ and $-3$ and no metal lines, as defined in Figure \ref{fig:synt}, with an extra offset of 0.05 for the no metal line. \label{fig:UMP}}
\end{figure}

Uncovering a sample of hundreds of extremely metal-poor stars is one of the main goals of the \emph{Pristine} survey. As detailed in the introduction, many scientific questions about the early Universe are hampered by the small number of known extremely metal-poor stars, but \emph{Pristine} provides the capability to efficiently find and study large samples of [Fe/H]$ < -3.0$ stars in a range of Galactic environments. Starting in 2016, the \emph{Pristine} collaboration began a dedicated spectroscopic follow-up program to gather spectra for the brightest extremely metal-poor candidates found in the \emph{Pristine} photometry (Venn et al., Aguado et al., in prep., Youakim et al., subm.). 

In Figure \ref{fig:UMP} we illustrate that \emph{Pristine} is well suited also to uncover the extremely rare ultra metal-poor stars (\FeH$ < -4.0$). Two additional MegaCam fields were strategically placed to cover the known ultra metal-poor stars SDSS J1742+2531 with [Fe/H] = $-4.80 \pm 0.07$ \citep[follow-up analysis with the Very Large Telescope instruments X-shooter and the Ultraviolet and Visual Echelle Spectrograph by][]{Caffau13b, Bonifacio15} and SDSS J183455+421328 with [Fe/H] = $-3.94 \pm 0.20$ \citep[follow-up analysis with Intermediate dispersion Spectrograph and Imaging System on the William Herschel Telescope instrument by][]{Aguado16}. These two stars are highlighted in Figure~\ref{fig:UMP} and their photometric uncertainties are represented by the error bars in the y-axis direction. It is obvious that the two ultra metal-poor stars indeed fall in the regime where we expect stars that have [Fe/H]$ < -3.0$ to be located, but still not above the no-metals line. They stand out significantly among most of the majority of metal-poor stars. Although, as expected, they do not stand out so significantly that \textit{Pristine} photometry can efficiently isolate them from the more numerous extremely metal-poor stars without follow up spectroscopy.

\subsection{Characterising the faint dwarf galaxies}

\begin{figure*}
\includegraphics[width=\linewidth]{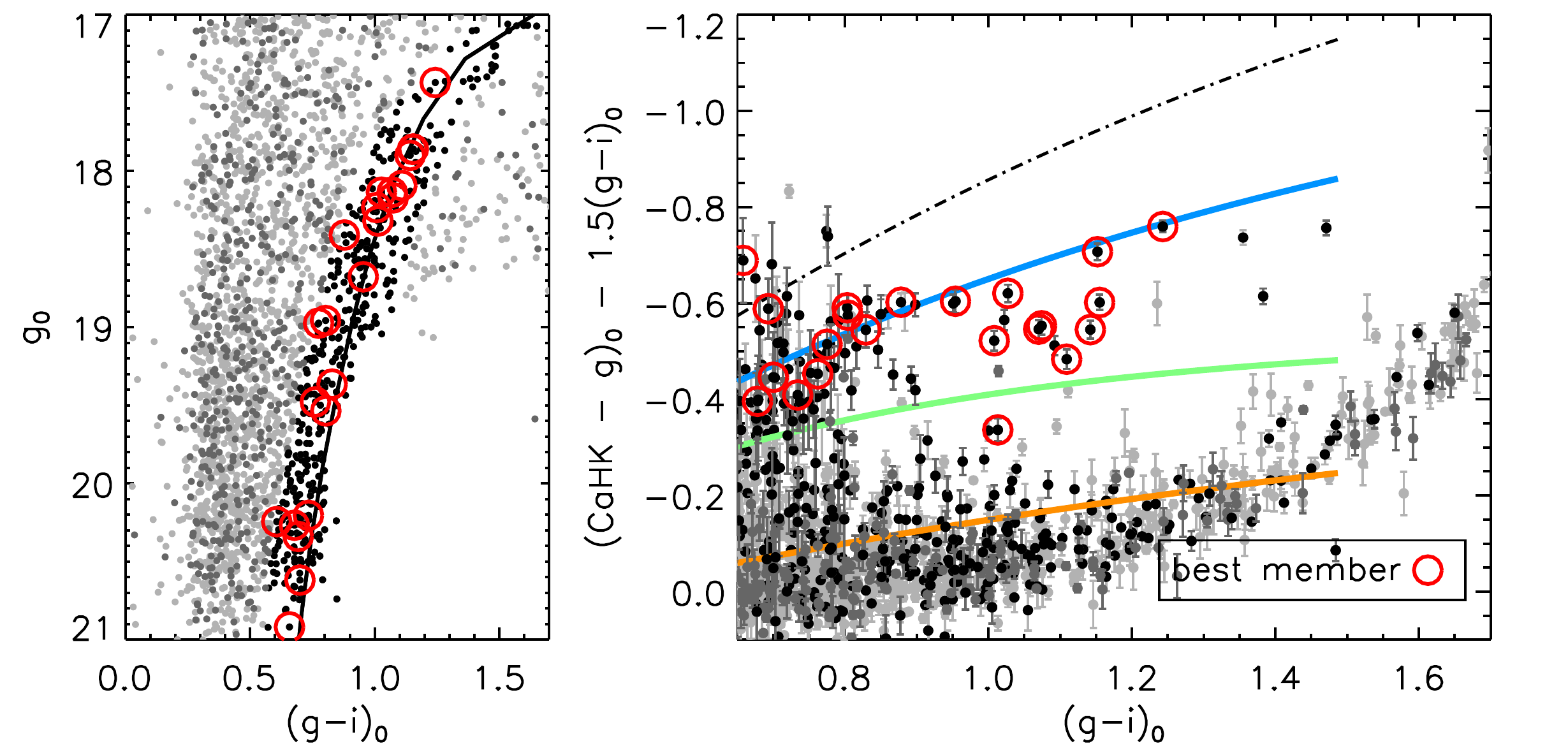}
\caption{Colour-magnitude diagram (left) and SDSS--\emph{Pristine} colour-colour plot (right) for \emph{Pristine} stars in a 4-deg$^2$ region centred on the Bo\"{o}tes I dwarf galaxy. Stars within the central degree are represented with a darker colour, stars along the stellar population sequence of M92 at the distance of Bo\"{o}tes I \citep[as taken from][]{Clem08} are represented in black. Spectroscopically followed-up stars from \citet{Koposov11} and \citet{Norris10} are circled red if they agree with the Bo\"{o}tes I systemic radial velocity and are labelled as the best member samples in either paper from subsequent analysis on the spectra in terms of stellar parameters such as stellar gravity. Additionally they are only plotted if they have $g-i > 0.6$. The shape of the colour-magnitude distribution on the left is affected by cuts made on the \emph{CaHK} uncertainties (see text and Figure \ref{fig:uncertainties}). The \emph{CaHK} observations clearly gives a handle on membership without spectroscopic information and allows for an efficient pre-screening of candidate member-stars out to large radii from the galaxy's center. \label{fig:Boo}}
\end{figure*}

The last couple of decades saw the discovery of numerous satellites in the Galactic halo. Satellite discoveries in the SDSS \citep[e.g.,][]{Belokurov07}, Pan-STARRS1 \citep[e.g.,][]{Laevens15}, and the Dark Energy Survey \citep[e.g.,][]{bechtol15} have provided us with powerful observational constraints in our cosmic backyard, especially to understand the faint end of galaxy formation in the preferred cosmological paradigm of $\Lambda$CDM \citep[e.g.,][]{Belokurov13}. However, good quality spectroscopy samples of at least tens of member stars per system are essential to both reach a good understanding of a system's dynamics and to accurately derive their chemical evolution history. One of the main difficulties that the community is encountering when studying these systems stems from their low contrast and the very expensive endeavour that studying their individual stars represents: it is very difficult to weed out the overwhelming population of foreground contaminants. Typically, only the central regions of these dwarf galaxies are dense enough to yield a good return on (observational time) investment when selected from broad-band photometry alone. Further out, one encounters a crippling fraction of contaminants. Yet, it should be noted that the outskirts of these systems are especially valuable if we wish to understand their dynamics and, from there, derive their masses. At present, most of the ``mass'' measurements assume dynamical equilibrium \citep[e.g.,][]{Martin07, Simon07, Simon11}, which is far from being proven for systems within a few tens of kpc. In fact, there are already hints that at least some systems have complex kinematics \citep{Ibata06,Collins16,Martin16}. Similarly, the chemical signature of these ``first galaxies'' is proving useful to model the first ages of galactic formation \citep[e.g.,][]{Ji16,Webster16}, but current models are based on, at most, a handful of bright stars per galaxy \citep{FrebelNorris15}.

Photometric metallicity information as offered by \emph{Pristine} helps to efficiently isolate candidate member stars out to the edges of the dwarf galaxies. From these clean samples, it is then possible to refine the structural properties of stellar systems, search for the presence of extra-tidal features at the edge, and build large and wide samples for a systematic spectroscopic follow-up.

\begin{figure*}
\includegraphics[width=\linewidth]{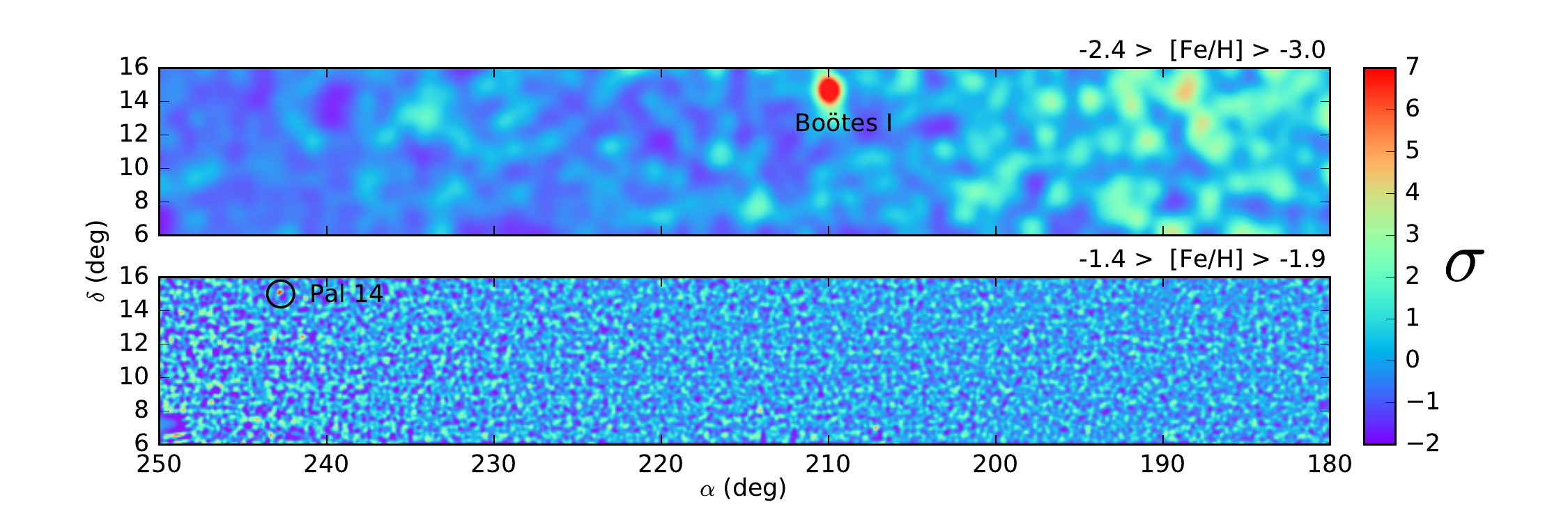}
\caption{Stellar density maps showing two known substructures of different metallicities and sizes contained within the \emph{Pristine} footprint for different $\FeH_\mathrm{Pristine}$ cuts : the Bo\"{o}tes I dwarf galaxy (top) and the globular cluster Pal 14 (bottom, circled). The parameter cuts to make these maps are based on literature values for the sizes and metallicities of these substructures (see text for details). The colour codes the significance of overdensities compared to the foreground/background stellar densities.\label{fig:BooSagPal}}
\end{figure*}

An illustration of the efficiency of this method can already be judged from relatively shallow \emph{Pristine} \emph{CaHK} observations of Bo\"{o}tes I, which lies within the survey footprint. Figure \ref{fig:Boo} shows the SDSS--\emph{Pristine} colour-colour plot for stars within a region of 4 deg$^{2}$ centred on Bo\"{o}tes I. The \emph{CaHK} uncertainty cuts are made such as to agree with the distance of the galaxy, we impose a cut of \emph{CaHK}$_{\rm err} < 0.05$ for the upper red giant branch (stars with $(g-i)_{0} > 0.8$) and \emph{CaHK}$_{\rm err} < 0.1$ for the full sample. Stars in this sample that are spectroscopically followed-up by either \citet{Norris10} or \citet{Koposov11} are circled in colour if the star is a very likely member of the Bo\"{o}tes I system, as determined by the radial velocity and stellar parameters derived from the spectra. Despite stemming from a short 100-second exposure, the CaHK photometry nicely identifies these best members and sets them, and other candidates, apart from the majority of foreground contamination.

One of the \emph{Pristine} science cases aims for a deeper coverage of all faint dwarf galaxies in the northern hemisphere to build clean samples of candidate member stars out to the edges of these systems for future spectroscopic studies.

\subsection{Mapping the metal-poor Galactic halo}

Within our Galactic halo we see a number of substructures, consisting of
dwarf galaxies, globular clusters and stellar streams \citep[e.g.,][]{Belokurov06a,Bernard16}. The survey area of \emph{Pristine} deliberately includes in its footprint a range of Galactic latitudes ($30 < b < 78\deg$) and some known stellar halo substructures. For instance, it includes part of the Sagittarius stellar stream and the Virgo overdensity, without however being overwhelmed by either \citep{Carlin12,Lokhorst16}. It is interesting to target the stellar streams to follow them in areas where they are less dense and overwhelmed by more metal-rich foreground stars. Additionally, studying the (low-)metallicity structure of these streams will help us better understand their process of destruction and their orbit around the Milky Way. Note that this mapping technique does not rely on spectroscopic follow-up and thus is applicable to faint stellar populations where spectroscopy is particularly costly to perform, such as, for instance, in external galaxies. Given the many existing deep broad-band imaging surveys, the addition of just one narrow-band filter to determine metallicities is relatively cheap and adds another dimension to Galactic substructure studies even in the era of the European Space Agency mission \textit{Gaia}.

Figure \ref{fig:BooSagPal} presents our ability to find and study substructures of different characteristic scales. We apply metallicity ranges and colour cuts taken from the literature and chosen to specifically select two known substructures contained within our survey footprint: the Bo\"{o}tes I dwarf galaxy \citep{Norris10} and the Palomar 14 (Pal 14) globular cluster \citep{Armandroff92}. The stellar density maps are created by placing stars into $1'$ bins and then smoothing by a Gaussian convolution kernel. A larger kernel (with $\sigma = 30'$) is used for Bo\"{o}tes I \citep{Belokurov06b}, while a smaller kernel (with $\sigma = 6'$) is used for Pal 14, to match the characteristic sizes of the substructures. Since our footprint extends towards the Galactic disc, there is a background density gradient which increases with right ascension. To remove this, we use a large convolution kernel ($60'$ for Bo\"{o}tes I and $20'$ for Pal 14) on the density map of all stars, and subtract it from the specific maps for each substructure. The top panel of Figure \ref{fig:BooSagPal} also highlights some substructures seen in the area to lower RA where our footprint is overlapping with the Sagittarius stellar stream. A further quantification of substructure at various metallicities can be compared to cosmological model expectations, and this will be the topic of future work.

\section{Conclusions}

In this work we present the \emph{Pristine} survey, built around MegaCam/CFHT observations conducted with a new narrow-band filter that covers the Ca H \& K doublet to efficiently isolate the Milky Way's most metal-poor stars. The current sky coverage extends over $\sim $1,000 deg$^2$, with \emph{CaHK} $\textrm{S/N} = 10$ at a depth of $g_0 \sim 21.0$. The \emph{Pristine} footprint includes a range of Galactic latitudes and several known substructures, such as the Sagittarius stream, the Virgo Overdensity, and several dwarf galaxies and globular clusters. Combined with broad-band SDSS photometry for object classification and temperature sensitivity, the Ca H \& K photometry is an excellent metallicity indicator. 

With a width of 100\AA, the CFHT CaHK filter is very narrow and its throughput shape is almost perfectly top-hat, resulting in a great metallicity sensitivity and little danger of ``leakage'' from other features such as strong molecular bands. We demonstrate that we achieve a calibration of the \emph{CaHK} magnitudes at the 0.02-magnitude level by relying on the stellar locus of red dwarf stars. This is more than sufficient for our science case.

Using SDSS/SEGUE spectra to calibrate the \emph{Pristine} photometric metallicities, we show in Figure \ref{fig:Segueclean} that the \emph{CaHK} observations have a high discriminatory power over the metallicity of a star. It is clear that over the temperature range from the turn-off ($(g-i)_0 \approx 0.2$) to the tip of the red giant branch ($(g-i)_0 \approx 1.5$) we are very sensitive to the metallicity of a given star. Over a large metallicity range from $\FeH = -0.5$ to $\FeH = -3.0$ we can derive photometric metallicities with uncertainties of only $\sim 0.2$ dex. We thereby open up a part of the metallicity distribution that was so far not reachable without spectroscopy.

We demonstrate that the \emph{Pristine} photometry can provide a very clean and complete sample of (extremely) metal-poor stars for spectroscopic follow-up (see Section \ref{sec:rates}) and show in Figure \ref{fig:UMP} that it is well positioned to uncover the very rare but extremely interesting ultra metal-poor stars that are possibly the most accessible messengers from the early Universe. Besides its great efficiency to select rare targets for spectroscopic follow-up studies, the photometry information alone allows for a careful (metallicity) mapping of substructures in the Galactic halo. We furthermore demonstrate how this new metallicity information can be used to very efficiently find new members of faint dwarf galaxies and metal-poor globular clusters, out to large radii, where foreground populations dominate.

In the near future, we expect the \emph{Pristine} survey to open a valuable window onto the regime of extremely-metal-poor stars and enable the construction of sizeable samples of stars with $\FeH < -3.0$ that can then be followed-up in more detail with spectroscopic campaigns.

\section*{Acknowledgements} 
We would like to thank the referee for constructive comments that helped improving this manuscript. We gratefully thank the CFHT staff for performing the observations in queue mode, for their reactivity in adapting the schedule, and for answering our questions during the data-reduction process. We thank Nina Hernitschek for granting us access to the catalogue of Pan-STARRS variability catalogue. ES and KY gratefully acknowledge funding by the Emmy Noether program from the Deutsche Forschungsgemeinschaft (DFG). NFM, RAI, and NL gratefully acknowledge funding from CNRS/INSU through the Programme National Galaxies et Cosmologie. ES and KY benefited from the International Space Science Institute (ISSI) in Bern, CH, thanks to the funding of the Team "The Formation and Evolution of the Galactic Halo". D.A. acknowledges the Spanish Ministry of Economy and Competitiveness (MINECO) for the financial support received in the form of a Severo-Ochoa PhD fellowship, within the Severo-Ochoa International Ph.D. Program. D.A., C.A.P., and J.I.G.H. also acknowledge the Spanish ministry project MINECO AYA2014-56359-P. J.I.G.H. acknowledges financial support from the Spanish Ministry of Economy and Competitiveness (MINECO) under the 2013 Ram\'{o}n y Cajal program MINECO RYC-2013-14875. 

Based on observations obtained with MegaPrime/MegaCam, a joint project of CFHT and CEA/DAPNIA, at the Canada-France-Hawaii Telescope (CFHT) which is operated by the National Research Council (NRC) of Canada, the Institut National des Science de l'Univers of the Centre National de la Recherche Scientifique (CNRS) of France, and the University of Hawaii.

The Pan-STARRS1 Surveys (PS1) have been made possible through contributions of the Institute for Astronomy, the University of Hawaii, the Pan-STARRS Project Office, the Max-Planck Society and its participating institutes, the Max Planck Institute for Astronomy, Heidelberg and the Max Planck Institute for Extraterrestrial Physics, Garching, The Johns Hopkins University, Durham University, the University of Edinburgh, Queen's University Belfast, the Harvard-Smithsonian Center for Astrophysics, the Las Cumbres Observatory Global Telescope Network Incorporated, the National Central University of Taiwan, the Space Telescope Science Institute, the National Aeronautics and Space Administration under Grant No. NNX08AR22G issued through the Planetary Science Division of the NASA Science Mission Directorate, the National Science Foundation under Grant No. AST-1238877, the University of Maryland, and Eotvos Lorand University (ELTE).

Funding for the SDSS and SDSS-II has been provided by the Alfred P. Sloan Foundation, the Participating Institutions, the National Science Foundation, the U.S. Department of Energy, the National Aeronautics and Space Administration, the Japanese Monbukagakusho, the Max Planck Society, and the Higher Education Funding Council for England. The SDSS Web Site is http://www.sdss.org/. The SDSS is managed by the Astrophysical Research Consortium for the Participating Institutions. The Participating Institutions are the American Museum of Natural History, Astrophysical Institute Potsdam, University of Basel, University of Cambridge, Case Western Reserve University, University of Chicago, Drexel University, Fermilab, the Institute for Advanced Study, the Japan Participation Group, Johns Hopkins University, the Joint Institute for Nuclear Astrophysics, the Kavli Institute for Particle Astrophysics and Cosmology, the Korean Scientist Group, the Chinese Academy of Sciences (LAMOST), Los Alamos National Laboratory, the Max-Planck-Institute for Astronomy (MPIA), the Max-Planck-Institute for Astrophysics (MPA), New Mexico State University, Ohio State University, University of Pittsburgh, University of Portsmouth, Princeton University, the United States Naval Observatory, and the University of Washington.

Funding for SDSS-III has been provided by the Alfred P. Sloan Foundation, the Participating Institutions, the National Science Foundation, and the U.S. Department of Energy Office of Science. The SDSS-III web site is http://www.sdss3.org/. SDSS-III is managed by the Astrophysical Research Consortium for the Participating Institutions of the SDSS-III Collaboration including the University of Arizona, the Brazilian Participation Group, Brookhaven National Laboratory, Carnegie Mellon University, University of Florida, the French Participation Group, the German Participation Group, Harvard University, the Instituto de Astrofisica de Canarias, the Michigan State/Notre Dame/JINA Participation Group, Johns Hopkins University, Lawrence Berkeley National Laboratory, Max Planck Institute for Astrophysics, Max Planck Institute for Extraterrestrial Physics, New Mexico State University, New York University, Ohio State University, Pennsylvania State University, University of Portsmouth, Princeton University, the Spanish Participation Group, University of Tokyo, University of Utah, Vanderbilt University, University of Virginia, University of Washington, and Yale University.

Funding for the Sloan Digital Sky Survey IV has been provided by the Alfred P. Sloan Foundation, the U.S. Department of Energy Office of Science, and the Participating Institutions. SDSS-IV acknowledges support and resources from the Center for High-Performance Computing at the University of Utah. The SDSS web site is www.sdss.org. SDSS-IV is managed by the Astrophysical Research Consortium for the Participating Institutions of the SDSS Collaboration including the Brazilian Participation Group, the Carnegie Institution for Science, Carnegie Mellon University, the Chilean Participation Group, the French Participation Group, Harvard-Smithsonian Center for Astrophysics, Instituto de Astrof\'isica de Canarias, The Johns Hopkins University, Kavli Institute for the Physics and Mathematics of the Universe (IPMU) / University of Tokyo, Lawrence Berkeley National Laboratory, Leibniz Institut f\"ur Astrophysik Potsdam (AIP), Max-Planck-Institut f\"ur Astronomie (MPIA Heidelberg), Max-Planck-Institut f\"ur Astrophysik (MPA Garching), Max-Planck-Institut f\"ur Extraterrestrische Physik (MPE), National Astronomical Observatories of China, New Mexico State University, New York University, University of Notre Dame, Observat\'ario Nacional / MCTI, The Ohio State University, Pennsylvania State University, Shanghai Astronomical Observatory, United Kingdom Participation Group,Universidad Nacional Aut\'onoma de M\'exico, University of Arizona, University of Colorado Boulder, University of Oxford, University of Portsmouth, University of Utah, University of Virginia, University of Washington, University of Wisconsin, Vanderbilt University, and Yale University.

LAMOST is operated and managed by the National Astronomical Observatories, Chinese Academy of Sciences.

\appendix

\section{Comparing SDSS stellar parameter pipeline product metallicities}\label{compareSDSS}

\begin{figure*}
\includegraphics[width=\linewidth]{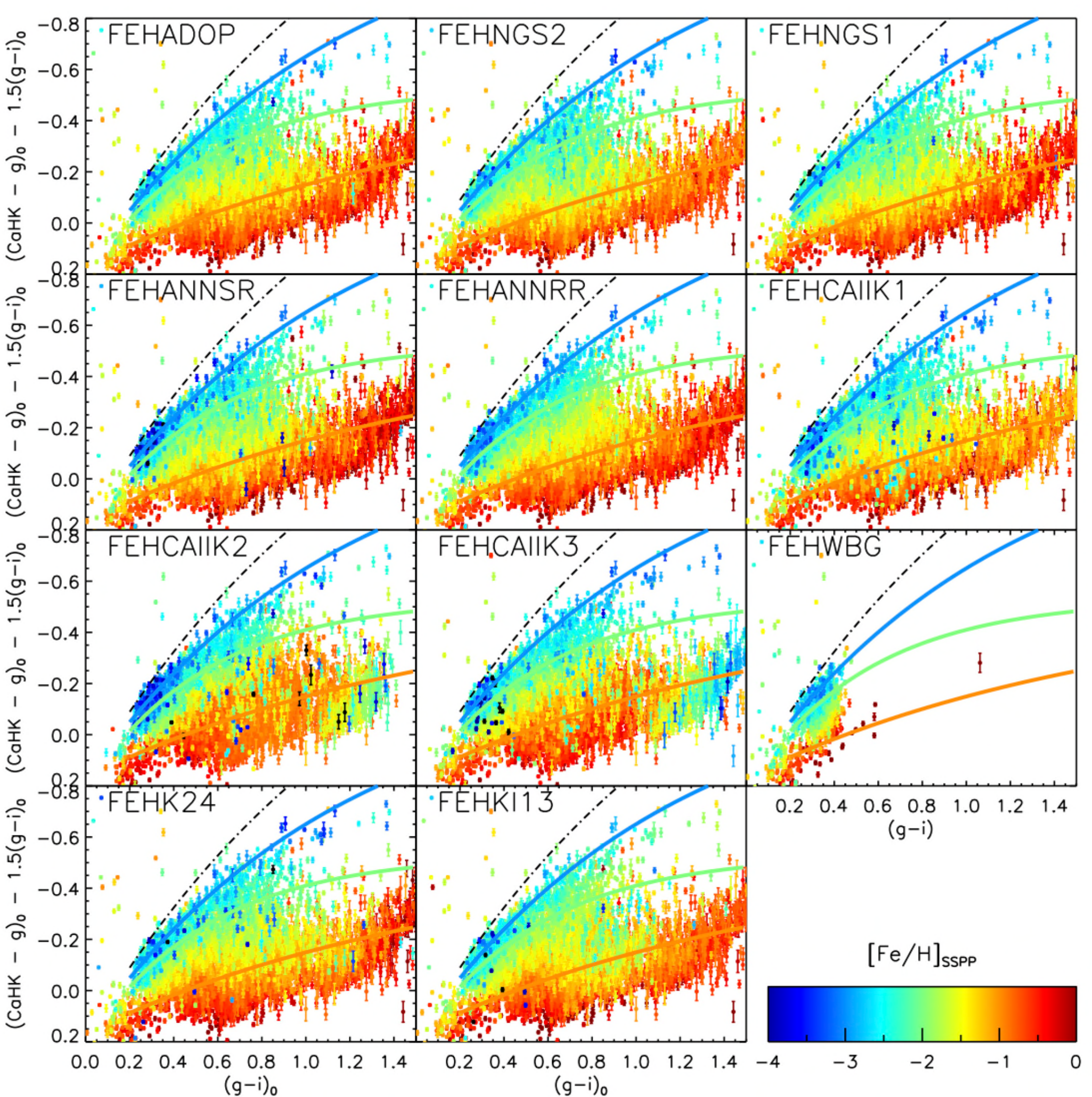}
\caption{Distribution of stars in common between SDSS/SEGUE and \emph{Pristine} in the \emph{CaHK}$_0$, $g_0$, and $i_0$ colour-colour space. Stars are colour-coded by their SDSS/SEGUE metallicity. The coloured lines are fits to synthetic spectra model predictions for stars with $\FeH = -1.0$, $-2.0$, $-3.0$, and with no metal lines (orange, green, blue, and black, respectively), as defined in Figure 3. Each panel corresponds to a different metallicity output from the SDSS stellar parameter pipeline SSPP. Quality cuts are described in the text of Section 3.}
\label{fig:SegueFeH}
\end{figure*}

The different panels of Figure \ref{fig:SegueFeH} show the various [Fe/H] estimates given by the SDSS collaboration from their stellar parameter pipeline (SSPP; see \citealt{Lee08} for an overview description and some of the methods and \citealt{Beers99}, \citealt{Wilhelm99}, \citealt{AllendePrieto06}, and \citealt{ReFiorentin07} for the original references for the FEHCAIIK2, FEHWBG, FEHK24, and FEHANNRR methods respectively). The KI13, K24, NGS1, and NGS2 are grid-matching-based methods. The WBG calibration method is particularly designed for hot stars \citep{Wilhelm99}. Both ANNSR and ANNRR are neural network approaches, respectively trained on synthetic or real spectra. CaIIK1 is determined from the 3850--4250 \AA\ region, whereas CaIIK2 and CaIIK3 are the [Fe/H] estimates based on the Ca II K line. 

For our calibration purposes, we choose not to use the SDSS adopted FEHADOP scale. This combination of all the SDSS [Fe/H] determinations might work well for the average properties of any stellar population, but can easily erase the signal in the tails of the distribution, particularly at the very metal-poor end, where not all methods will deliver reliable results. Focussing instead on the ordering of the colour-coding in the colour-colour space of Figure~\ref{fig:SegueFeH} and minimising the scatter in this plane, we find that the FEHANNRR method agrees best with our colour-colour map at the low-metallicity end. This result is quantified in Table \ref{tab:stats}. Here, we compare the results from each of the SDSS metallicity scales to the predictions for \FeH$ = -1,-2,$ and $-3$ from the synthetic spectra as illustrated in Figure 3. We use two different comparisons. First, we calculate the offset from \FeH$_\mathrm{syn}$ to \FeH$_\mathrm{SSPP}$ for each star that has a vertical distance less than 0.02 from each of the synthetic spectra line fits. Second, we calculate the vertical distance for each star that has \FeH$_\mathrm{SSPP} - \FeH_\mathrm{syn} < 0.2$ to its corresponding synthetic spectra line fit. Median values and standard deviations for these distributions are given in Table \ref{tab:stats}. While the first set of statistics mainly check the uniformity of the metallicity space around the expected \FeH$ = -1,-2,$ and $-3$ loci, the second test quantifies if there are many stars scattered over the rest of the colour-colour space that have been given this metallicity value in the corresponding SSPP metallicity scale. Interestingly, for instance, any of the metallicity estimates based on the Ca H \& K region (FEHCAIIK1, FEHCAIIK2, and FEHCAIIK3) correlate rather poorly with the \emph{Pristine} photometry in the sense that they give very low metallicities to stars that live in places of the colour-colour space where high metallicities are expected. We presume this is due to the typically much lower SNR in this region of the spectra. When compared to the synthetic spectra expectations it can also be seen from Table \ref{tab:stats} that FEHANNRR behaves much better at $-3$ than FEHADOP at the lowest metallicities. Not only are the [Fe/H] values closer to $-3$ for stars around the synthetic $-3$ line (the median difference is 0.54 compared to 0.79) but, additionally, the stars that are flagged as $\sim -3$ by the pipeline are more clustered around this line (the standard deviation is 0.038 instead of 0.067). Also for stars around the $-2$ synthetic line, the values from FEHANNRR are closer to $-2$ (a median offset of 0.018 instead of 0.157). At the same time the performance at higher metallicities is comparable. 

\begin{table*}
	\centering
	\caption{Statistics to compare the ordering and scatter of metallicity information in the colour-colour plane shown in Figure \ref{fig:SegueFeH} when compared to expectations from synthetic spectra as presented in Figure 3. For \FeH$ = -1,-2,$ or $-3$, stars close to fits to the synthetic spectral lines are selected and their metallicity offsets evaluated (columns 3 \& 4 in this table give the median and standard deviation values). Additionally, for stars with $\FeH_\mathrm{SSPP}$ metallicities close to these values we consider their vertical offsets to these fit lines (columns 5 \& 6). \label{tab:stats}}
	\begin{tabular}{cccccc} % four columns, alignment for each
		\hline
		\hline
		Method & $\FeH_\mathrm{syn}$ & median($\FeH_\mathrm{syn}$ & $\sigma$($\FeH_\mathrm{syn}$ & median(y$_\mathrm{syn}$ & $\sigma$(y$_\mathrm{syn}$ \\
		& &\ $- \FeH_\mathrm{SSPP}$) & \ $- \FeH_\mathrm{SSPP}$) & \ $-$y$_\mathrm{SSPP}$) & \ $-$y$_\mathrm{SSPP}$) \\
		\hline
		
		\hline
FEHADOP & $-3$ & 0.794 & 0.344 & 0.000 & 0.067 \\ 
 & $-2$ & 0.157 & 0.306 & -0.010 & 0.050 \\ 
 & $-1$ & 0.226 & 0.284 & -0.033 & 0.082 \\ 
\hline
FEHNGS2 & $-3$ & 0.895 & 0.390 & 0.029 & 0.070 \\ 
 & $-2$ & 0.210 & 0.392 & -0.008 & 0.065 \\ 
 & $-1$ & 0.103 & 0.315 & -0.014 & 0.087 \\ 
\hline
FEHNGS1 & $-3$ & 0.905 & 0.374 & 0.027 & 0.129 \\ 
 & $-2$ & 0.227 & 0.324 & -0.012 & 0.058 \\ 
 & $-1$ & 0.205 & 0.325 & -0.030 & 0.089 \\ 
\hline
FEHANNSR & $-3$ & 0.695 & 0.538 & 0.005 & 0.072 \\ 
 & $-2$ & 0.142 & 0.410 & -0.002 & 0.061 \\ 
 & $-1$ & 0.153 & 0.344 & -0.015 & 0.088 \\ 
\hline
FEHANNRR & $-3$ & 0.543 & 0.370 & 0.005 & 0.038 \\ 
 & $-2$ & 0.018 & 0.346 & -0.001 & 0.055 \\ 
 & $-1$ & 0.251 & 0.280 & -0.035 & 0.081 \\ 
\hline
FEHCAIIK1 & $-3$ & 0.719 & 0.429 & 0.033 & 0.093 \\ 
 & $-2$ & 0.069 & 0.380 & 0.000 & 0.062 \\ 
 & $-1$ & 0.164 & 0.368 & -0.028 & 0.089 \\ 
\hline
FEHCAIIK2 & $-3$ & 0.342 & 0.457 & 0.015 & 0.056 \\ 
 & $-2$ & 0.046 & 0.459 & 0.003 & 0.096 \\ 
 & $-1$ & 0.269 & 0.397 & -0.046 & 0.093 \\ 
\hline
FEHCAIIK3 & $-3$ & 0.733 & 0.459 & 0.044 & 0.243 \\ 
 & $-2$ & 0.198 & 0.341 & -0.005 & 0.110 \\ 
 & $-1$ & 0.358 & 0.522 & -0.064 & 0.082 \\ 
\hline
FEHWBG & $-3$ & 0.639 & 0.381 & 0.000 & 0.029 \\ 
 & $-2$ & -0.024 & 0.298 & 0.011 & 0.056 \\ 
 & $-1$ & 0.239 & 0.540 & -0.067 & 0.085 \\ 
\hline
FEHK24 & $-3$ & 0.746 & 0.392 & 0.023 & 0.111 \\ 
 & $-2$ & 0.193 & 0.363 & -0.014 & 0.059 \\ 
 & $-1$ & 0.250 & 0.325 & -0.040 & 0.080 \\ 
\hline
FEHKI13 & $-3$ & 0.851 & 0.339 & -0.011 & 0.131 \\ 
 & $-2$ & 0.204 & 0.324 & -0.016 & 0.070 \\ 
 & $-1$ & 0.246 & 0.342 & -0.044 & 0.092 \\ 
\hline
\hline		
	\end{tabular}
\end{table*}

\bibliography{references}

\end{document}